# A Novel Unified Expression for the Capacity and Bit Error Probability of Wireless Communication Systems over Generalized Fading Channels

Ferkan Yilmaz and Mohamed-Slim Alouini

Electrical Engineering Program, Division of Physical Sciences and Engineering,

King Abdullah University of Science and Technology (KAUST),

Thuwal, Mekkah Province, Saudi Arabia.

Email(s): {*ferkan.yilmaz, slim.alouini*}*@kaust.edu.sa*

**Abstract**

Analysis of the average binary error probabilities (ABEP) and average capacity (AC) of wireless communications systems over generalized fading channels have been considered separately in the past. This paper introduces a novel moment generating function (MGF)-based *unified expression* for the ABEP and AC of single and multiple link communication with maximal ratio combining. In addition, this paper proposes the hyper-Fox's H fading model as a unified fading distribution of a majority of the well-known generalized fading models. As such, we offer a generic unified performance expression that can be easily calculated and that is applicable to a wide variety of fading scenarios. The mathematical formalism is illustrated with some selected numerical examples that validate the correctness of our newly derived results.

## I. INTRODUCTION

The average binary error probabilities (ABEP) and average capacity (AC) are important performance metrics of wireless communication systems operating over fading channels. As such, considerable efforts have been devoted so far to develop analytical tools/frameworks to evaluate these two performance metrics [1, and the references therein]. However, and to the best authors' knowledge, these tools/frameworks were developed separately and the computation of these two performance metrics was viewed as two independent problems. For instance, based on Craig's representation of the complementary error function, a unified moment generating function (MGF)-based approach was developed to compute the ABEP of a wide variety of modulation techniques over generalized fading [1, and the references therein]. More recently, other MGF-based approaches [2]–[4] were also proposed for the capacity calculation of wireless channels subject to generalized fading.

In contrast, this paper presents a novel MGF-based *unified expression* for the exact evaluation of both the ABEP and AC of single an multiple links generalized faded channels. The paper introduces also the hyper-Fox's H distribution as a versatile fading model including a variety of well-known models as special cases. With these two unifying frameworks at hand, we propose new generic expressions for the ABEP and AC with and without diversity reception. We also present some selected numerical examples to validate our newly derived results.

The remainder of this paper is organized as follows. In Section II, a unified performance measure analysis of diversity receivers over additive white Gaussian noise (AWGN) channels is introduced and some key results are presented. In Section III, new results for single link and multiple link reception are presented and applied to the newly proposed unifying hyper-Fox's H fading model. Finally, the main results are summarized and some conclusions are drawn in the last section.

## II. Unified Conditional Performance Expression

A compact form for the conditional bit error probability (BEP) $P_{BEP}(\gamma_{end})$ for a certain value of instantaneous SNR $\gamma_{end}$ for different binary modulations was proposed by Wojnar in [5, Eq. (13)] as

$$P_{BEP}(\gamma_{end}) = \frac{\Gamma(b, a\gamma_{end})}{2\Gamma(b)}, \quad a, b \in \left\{1, \frac{1}{2}\right\}, \tag{1}$$

where $a$ depends on the type of modulation ($\frac{1}{2}$ for orthogonal frequency shift keying (FSK), $1$ for antipodal phase shift keying (PSK)), $b$ depends on the type of detection ($\frac{1}{2}$ for coherent, $1$ for non-coherent), and $\Gamma(\cdot, \cdot)$ denotes the complementary incomplete Gamma function [6, Eq. (6.5.3)]. In the following theorem, we introduce an alternative representation of (1) using the incomplete beta function.

**Theorem 1** (Unified BEP Expression Using the Incomplete Beta Function). *An alternative representation*

*of the compact form of the conditional bit error probability $P_{BEP}(\gamma_{end})$ represented in* (1) *is given by*

$$P_{BEP}(\gamma_{end}) = \frac{1}{2} - \frac{\exp(-\mathrm{i}\pi b)}{2\Gamma(b)} \lim_{d \to \infty} d^b B\left(-\frac{a}{d}\gamma_{end}; b, 1-d\right), \tag{2}$$

*where the parameters $a$ and $b$ depend on the particular form of the modulation and detection as mentioned above (see Table I), and* $\mathrm{i} = \sqrt{-1}$ *is the imaginary unit with the property* $\mathrm{i}^2 = -1$*, and* $B(z; a, b) = \int_0^z u^{a-1}(1-u)^{b-1} du$ *is the incomplete beta function* [6, Eq. (6.6.1)].

*Proof:* See Appendix A. ∎

In addition to the BEP performance measure, there exists another important performance measure commonly used in the literature, which is known as the conditional capacity. Explicitly, the conditional capacity is a measure of how much error-free information can be transmitted and received through the channel. The conditional normalized channel capacity $P_C(\gamma_{end})$ in nats/s/Hz for a certain value of instantaneous SNR $\gamma_{end}$ at the output of the receiver is well-known to be given by

$$P_C(\gamma_{end}) = \log(1 + a\,\gamma_{end}), \quad \text{nats/s/Hz}, \tag{3}$$

where $a \in \mathbb{R}^+$ represents the transmission power, and $\log(\cdot)$ is the natural logarithm (i.e., the logarithm to the base $e$) [6, Eq. (4.1.1)]. We introduce in the following theorem a new alternative incomplete beta function-based representation of (3).

**Theorem 2** (Capacity Expression Using the Incomplete Beta Function). *An alternative representation of the conditional capacity $P_C(\gamma_{end})$ given in* (3) *is given by*

$$P_C(\gamma_{end}) = -B(-a\gamma_{end}; 1, 0). \tag{4}$$

*Proof:* See Appendix B. ∎

To the best of authors' knowledge, both (2) and (4) are not available in the literature. More importantly and using these two incomplete beta representations of the BEP and channel capacity measures, one can readily give, as shown in the following corollary, a unified performance expression whose special cases include the BEP and channel capacity.

**Corollary 1** (Unified Performance Expression Using Incomplete Beta Function). *A compact and unified form of the conditional performance measure $P_{UP}(\gamma_{end})$ (which include both the conditional BEP and* *the conditional channel capacity) is given by*

$$P_{UP}(\gamma_{end}) = \alpha + \beta \frac{\exp(-i\pi b) d^b}{2\Gamma(b)} B\left(-\frac{a}{d}\gamma_{end}; b, 1-d\right). \tag{5}$$

*which reduces to* (1) *for* $\alpha = 1$, $\beta = -1$ *and* $d \to \infty$, *and which reduces to* (3) *for* $\alpha = 0$, $\beta = 2$, $a = 1$, $b = 1$ *and* $d = 1$.

*Proof:* Based on Theorem 1 and Theorem 2, the proof is obvious. ∎

In the limit as $d \to \infty$, (5) might produce approximate results due to numerical computation limits of standard mathematical software packages. In this context, using both [7, Eq. (7.2.2/13)] and [7, Eq. (7.3.1/28)], the unified expression given by Corollary 1 can be represented without limit as shown in the following corollary.

**Corollary 2** (Unified Performance Expression Using the Hypergeometric Function). *The unified performance measure* $P_{UP}(\gamma_{end})$ *can also be represented as*

$$P_{UP}(\gamma_{end}) = 1 - \frac{n}{2}\left\{1 + (-1)^n \frac{(a\gamma_{end})^b}{\Gamma(b+1)} {}_nF_1\left[\Lambda_b^{(n)}; b+1; -a\gamma_{end}\right]\right\}, \tag{6}$$

*which reduces to* (1) *for* $n = 1$, *and which reduces to* (3) *for* $a = 1, b = 1$ *and* $n = 2$. *Moreover, in* (6), *the coefficient set* $\Lambda_a^{(m)}$ *is defined as*

$$\Lambda_a^{(n)} \equiv \overbrace{a, a, \ldots, a}^{n \text{ times}}, \tag{7}$$

*and* ${}_pF_q[\cdot;\cdot;\cdot]$ *denotes the generalized hypergeometric function [7, Eq. (7.2.3/1)].*

*Proof:* Using both [7, Eq. (7.2.2/13)] and [7, Eq. (7.3.1/28)], the proof is obvious. ∎

Note that, in addition to the hypergeometric function representation given by Corollary 2, the unified expression can also be given in terms of other special functions such as the MacRobert's E function and the Meijer's G function, as shown in the following corollaries. These expressions are useful since they will facilitate in Section III the unified analysis of BEP and capacity in generalized fading environments.

**Corollary 3** (Unified Performance Expression Using the MacRobert's E Function). *The unified performance measure* $P_{UP}(\gamma_{end})$ *can also be written as*

$$P_{UP}(\gamma_{end}) = 1 - \frac{n}{2}\left\{1 - (-1)^n \frac{(a\gamma_{end})^b}{\Gamma(b)} E\left[\Lambda_b^{(n)}; b+1; \frac{1}{a\gamma_{end}}\right]\right\} \tag{8}$$

*where* $E[\cdot;\cdot;\cdot]$ *is the MacRobert's E function [8, Sec. (9.4)], [7, Sec. (2.23)]. Moreover,* (8) *reduces to*

(1) *for $n = 1$, and it reduces to* (3) *for $a = 1, b = 1$ and $n = 2$.*

*Proof:* Using the generalized hypergeometric function representation of the MacRobert's E function [7, Sec. (2.23)], i.e., substituting the following equality

$$E[a_1, a_2, \ldots, a_r; b_1, b_2, \ldots, b_q; az] = \frac{\prod_{r=1}^{p} \Gamma(a_r)}{\prod_{r=1}^{q} \Gamma(b_r)} \, _pF_q\left[a_1, a_2, \ldots, a_r; b_1, b_2, \ldots, b_q; -\frac{1}{az}\right] \quad (9)$$

into (6) results in (8), which proves Corollary 3. ∎

**Corollary 4** (Unified Performance Expression Using the Meijer's G Function). *The unified performance measure $P_{UP}(\gamma_{end})$ has this additional Meijer's G function-based representation given by*

$$P_{UP}(\gamma_{end}) = 1 - \frac{n}{2}\left\{1 - \frac{(-1)^n}{\Gamma(b)} G_{n,2}^{1,n}\left[a\gamma_{end} \,\bigg|\, \begin{matrix}\Lambda_1^{(n)} \\ b, 0\end{matrix}\right]\right\}, \quad (10)$$

*where $G_{p,q}^{m,n}[\cdot]$ is the Meijer's G function [7, Eq. (8.3.22)]. Furthermore, (10) reduces to (1) for $n = 1$, and it also reduces to (3) for $a = 1, b = 1$ and $n = 2$.*

*Proof:* Using the relation between MacRobert's E and Meijer's G function [7, Eq. (8.4.51/8)], we can write

$$E\left[\Lambda_b^{(n)}; b + 1; \frac{1}{a\gamma_{end}}\right] = G_{n,2}^{1,n}\left[a\gamma_{end} \,\bigg|\, \begin{matrix}\Lambda_{1-b}^{(n)} \\ 0, -b\end{matrix}\right]. \quad (11)$$

Then, substituting (11) into (6) results in (10), which proves Corollary 4. ∎

It is worth to mention that the Meijer's G function is an special function defined by the Mellin-Barnes type integral which contains products and quotients of the Euler gamma functions, and it is as such considered as a generalization of hypergeometric functions and other special functions such as exponential, bessel, logarithm, sine / cosine integral functions. Therefore, the PDF of some well-known fading distributions can be represented in terms of Meijer's G function. In this context, referring to [7, Eq. (2.24.1/1)], the representation in Corollary 4 is more useful than the representations given in Corollaries 1, 2 and 3 from numerical simplicity and computation start-points.

In the following section, we use our new representations in the analysis of the BEP and capacity measures in generalized fading environments.

## III. UNIFIED PERFORMANCE EXPRESSION OVER FADING CHANNELS

Over the past four decades, the ABEP and the AC measures have been considered as two different problems, and many different solutions ranging from bounds to approximations, integral expressions, and closed-form formulas have been presented for these two different measures for a variety of modulation schemes, diversity combining techniques, and fading distributions. At this point, we would also like to highlight again that these two different performance measures can be compactly combined as mentioned in the previous sections and can also be considered as a single problem. As such, for a certain nonnegative distribution of the instantaneous SNR $\gamma_{end}$ at the output of the receiver (i.e., $\gamma_{end}$ is distributed over $(0, \infty)$ according to $p_{\gamma_{end}}(\gamma)$ which is the probability density function (PDF) of $\gamma_{end}$), the average unified performance (AUP) expression is completely represented as

$$P_{AUP} = \mathbb{E}[P_{UP}(\gamma_{end})] = \int_0^\infty P_{UP}(\gamma) \, p_{\gamma_{end}(\gamma)} d\gamma, \quad (12)$$

where $\mathbb{E}[\cdot]$ denotes the expectation operator, and as proposed in Section II, $P_{UP}(\gamma)$ is simply the conditional unified performance expression.

## A. Single Link Reception

We consider an optimum receiver employing binary modulation and operating in a slow non-selective generalized fading environment corrupted by AWGN noise. In such a case, $\gamma_{end} = \gamma_\ell$ is the instantaneous SNR at the output of the receiver, and it is distributed over $(0, \infty)$ according to the PDF $p_{\gamma_\ell}(\gamma)$. Referring to (12), the AUP expression can thus written as

$$P_{AUP} = 1 - \frac{n}{2}\left\{1 - \frac{(-1)^n}{\Gamma(b)}\int_0^\infty G_{n,2}^{1,n}\left[a\gamma \,\middle|\, \begin{matrix}\Lambda_1^{(n)}\\b,0\end{matrix}\right] p_{\gamma_\ell}(\gamma)d\gamma\right\}. \quad (13)$$

The PDF of the distribution of the fading is a nonnegative function (i.e., $p_{\gamma_\ell}(\gamma) \geq 0$ for $\gamma \geq 0$) and a nonnegative function can be expressed in terms of Meijer's G or Fox's H function using integral transforms theory. In other words, the PDF of a variety of statistical envelope distributions, such as Rayleigh, Rician, exponential, Nakagami-*m*, Weibull, generalized Nakagami-*m*, lognormal, K-distribution, generalized-K, etc., can be expressed in terms of either Meijer's G or Fox's H function [9]. As a consequence of that, (13) can be readily reduced to a closed-form solution by exploiting [7, Eq. (2.24.1/1)]. In this context, the Fox's H channel fading model is introduced in the following definition.

**Definition 1** (Hyper-Fox's H Fading Channel)**.** *Let $\gamma_\ell$ be a hyper-Fox's H fading distribution representing the instantaneous SNR $\gamma_\ell$ at the output of the single-link receiver, which follows a probability law in* *general such that its hyper-Fox's H PDF is given by*

$$p_{\gamma_\ell}(\gamma) = \sum_{n_\ell=1}^{K_\ell} \eta_{n_\ell} \operatorname{H}_{P_{n_\ell},Q_{n_\ell}}^{M_{n_\ell},N_{n_\ell}} \left[ c_{n_\ell} \gamma \;\middle|\; \begin{array}{c} (\vartheta_{n_\ell j}, \theta_{n_\ell j})_{j=1}^{P_{n_\ell}} \\ (\varphi_{n_\ell j}, \phi_{n_\ell j})_{j=1}^{Q_{n_\ell}} \end{array} \right], \quad \gamma > 0, \tag{14}$$

*with the conditions*

$$\max_{j=1,2,\ldots,M_{n_\ell}} \left\{ -\frac{\varphi_{n_\ell j}}{\phi_{n_\ell j}} \right\} < \mathfrak{S}_{n_\ell} < \min_{j=1,2,\ldots,N_{n_\ell}} \left\{ \frac{1-\vartheta_{n_\ell j}}{\theta_{n_\ell j}} \right\}, \quad n_\ell \in \{1,2,\ldots,K_\ell\}, \tag{15}$$

*such that* $\mathfrak{S} = \mathfrak{S}_1 \cap \mathfrak{S}_2 \cap \ldots \cap \mathfrak{S}_{K_\ell} \neq \emptyset$, *where both* $\{(\varphi_{n_\ell j}, \phi_{n_\ell j})_{j=1}^{Q_{n_\ell}}\}$ *and* $\{(\vartheta_{n_\ell j}, \theta_{n_\ell j})_{j=1}^{P_{n_\ell}}\}$ *coefficient sets are such parameters that they support the conditions given by* [7, Section 8.3.1], *and where* $\eta_{n_\ell} \in \mathbb{R}$ *and* $c_{n_\ell} \in \mathbb{R}^+$ *are such two parameters that (14) certainly supports* $\int_0^\infty p_{\gamma_\ell}(\gamma)\,d\gamma = 1$. *In (14),* $\operatorname{H}_{p,q}^{m,n}[\cdot]$ *denotes the Fox's H function* [7, Eq. (8.3.1/1)][1],[2].

It is worth to notice that the PDF of most non-negative distributions can be compactly expressed or accurately approximated in the form of (14). For example, as seen in the first column of Tables II,III,V and V, the PDFs commonly used in the literature for modeling of instantaneous SNR distribution $\gamma_{end}$ are listed in the form of hyper-Fox's fading distribution. *From this perspective, where hyper-Fox's H distribution provides an unified framework on modeling of fading distribution, the unified and generalized result regarding the AUP expression of single-link reception can be obtained* by means of substituting (14) into (13) and using [11, Theorem 2.9] with [7, Eq. (8.3.2/21)] such that the AUP expression over hyper-Fox's H fading channel is given by

$$P_{AUP} = 1 - \frac{n}{2} \left\{ 1 - \frac{(-1)^n}{\Gamma(b)} \sum_{n=1}^{K_\ell} \frac{\eta_{n_\ell}}{c_{n_\ell}} \operatorname{H}_{P_{n_\ell}+2,Q_{n_\ell}+n}^{M_{n_\ell}+n,N_{n_\ell}+1} \left[ \frac{c_{n_\ell}}{a} \;\middle|\; \begin{array}{c} (1-b,1), (\vartheta_{n_\ell j}+\theta_{n_\ell j}, \theta_{n_\ell j})_{j=1}^{P_{n_\ell}}, (1,1) \\ \Lambda_{((0,1))}^{(n)}, (\varphi_{n_\ell j}+\phi_{n_\ell j}, \phi_{n_\ell j})_{j=1}^{Q_{n_\ell}} \end{array} \right] \right\} \tag{16}$$

with the condition $\mathfrak{P} \cap \{\mathfrak{S}_1 \cap \mathfrak{S}_2 \cap \ldots \cap \mathfrak{S}_{K_\ell}\} \neq \emptyset$, where $0 < \mathfrak{P} < b$ and the convergence region set $\{\mathfrak{S}_{n_\ell}\}$ is defined in (15).

Let us consider some special cases of the hyper-Fox's H fading model in order to check analytical simplicity, accuracy and correctness of the AUP expression given by (13).

**Example 1** (Unified Performance Measure in Generalized Nakagami-*m* (GNM) Fading Channels)**.** In GNM fading channels, the distribution of the instantaneous SNR $\gamma_\ell$ follows a generalized Gamma PDF

---

[1] For more information about the Fox's H function, the readers are referred to [10], [11]

[2] Using [7, Eq. (8.3.22)], the Fox's H function can be represented in terms of the Meijer's G function [7, Eq. (8.2.1)] which is a built-in function in the most popular mathematical software packages such as MATHEMATICA®.

given by

$$p_{\gamma_\ell}(\gamma) = \frac{\xi_\ell}{\Gamma(m_\ell)}\left(\frac{\beta_\ell}{\bar{\gamma}_\ell}\right)^{\xi_\ell m_\ell}\gamma^{\xi_\ell m_\ell - 1}e^{-\left(\frac{\beta_\ell}{\bar{\gamma}_\ell}\right)^{\xi_\ell}\gamma^{\xi_\ell}} \qquad (17)$$

for $0 \geq \gamma < \infty$, where the parameters $m_\ell \geq 1/2$, $\xi_\ell > 0$ and $\bar{\gamma}_\ell > 0$ are the fading figure, the shaping parameter and the local mean power of the $\ell$th GNM distribution, and $\beta_\ell = \Gamma(m_\ell + 1/\xi_\ell)/\Gamma(m_\ell)$. It may be useful to notice that the special or limiting cases of the GNM distribution are well-known in the literature as the Rayleigh ($m_\ell = 1, \xi_\ell = 1$), exponential ($m_\ell = 1, \xi_\ell = 1/2$), Half-Normal ($m_\ell = 1/2, \xi_\ell = 1$), Nakagami-$m$ ($\xi_\ell = 1$), Gamma ($\xi_\ell = 1/2$, Weibull ($m_\ell = 1$), lognormal ($m_\ell \to \infty, \xi_\ell \to 0$), and AWGN ($m_\ell \to \infty, \xi_\ell = 1$). Referring to Definition 1 in order to obtain the AUP expression $P_{AUP}$ for generalized Gamma fading channels by means of using (16), one can readily represents the PDF $p_{\gamma_\ell}(\gamma)$ of the generalized Gamma distribution in terms of the hyper-Fox's H distribution by means of using [11, Eq. (2.9.4)], i.e., using the Fox's H representation given in the first row of Table V

$$p_{\gamma_\ell}(\gamma) = \frac{\beta_\ell}{\bar{\gamma}_\ell \Gamma(m_\ell)} H_{0,1}^{1,0}\left[\frac{\beta_\ell}{\bar{\gamma}_\ell}\gamma \,\middle|\, \begin{matrix}\text{---}\\(m_\ell - 1/\xi_\ell, 1/\xi_\ell)\end{matrix}\right] \qquad (18)$$

where —— means that the coefficients are absent. Next, expressing (18) in terms of (14) with $K = 1$, and then substituting (18) into (16) (i.e., by means of mapping the parameters of two PDF model (14) and (18), and then utilizing (16)), one can readily obtain the AUP expression for GNM fading channels as

$$P_{AUP} = 1 - \frac{n}{2}\left\{1 - \frac{(-1)^n}{\Gamma(b)\Gamma(m_\ell)} H_{2,n+1}^{n+1,1}\left[\frac{\beta_\ell}{a\bar{\gamma}_\ell}\,\middle|\,\begin{matrix}(1-b,1),(1,1)\\ \Lambda_{(0,1)}^{(n)},(m_\ell,1/\xi_\ell)\end{matrix}\right]\right\} \qquad (19)$$

Here, it might be useful to notice that, in addition to this AUP expression for GNM fading channels, Table I offers simplified expressions for the AUP of a variety of commonly used fading channels in order to facilitate for the readers the use of our unified BEP and AC results. In order to check the validity and completeness of (19), substituting $n = 1$ results in the ABEP of signal transmission over GNM fading channels as expected, that is

$$P_{ABEP} = \frac{1}{2}\left\{1 - \frac{1}{\Gamma(b)\Gamma(m_\ell)} H_{2,2}^{2,1}\left[\frac{\beta_\ell}{a\bar{\gamma}_\ell}\,\middle|\,\begin{matrix}(1-b,1),(1,1)\\(0,1),(m_\ell,1/\xi_\ell)\end{matrix}\right]\right\}. \qquad (20)$$

After performing some algebraic manipulations either by means of using [11, Property 2.11] or by recognizing that the cumulative distribution function (CDF) of a Fox's H distribution can be written in two different forms given in [12, Eq. (4.17)] and [12, Eq. (4.19)], (20) can be simplified to

$$P_{ABEP} = \frac{1}{2\Gamma(b)\Gamma(m_\ell)} H_{2,2}^{1,2}\left[\frac{\beta_\ell}{a\bar{\gamma}_\ell} \left| \begin{array}{c} (1,1),(1-b,1) \\ (m_\ell, 1/\xi_\ell),(0,1) \end{array}\right.\right]. \qquad (21)$$

in perfect agreement with [13, Eq. (10)]. Note that in the special case of Nakagami-$m$ fading channels, substituting $\xi_\ell = 1$, it is straight forward to show that (20) reduces to [5, Eq. (17)] by means of some algebraic manipulations using [7, Eq.(8.4.49/13)] and [7, Eq.(7.3.1/28)] together. In addition, for Rayleigh fading channel (i.e., $m_\ell = \xi_\ell = 1$), (20) simplifies to $P_{ABEP} = \frac{1}{2}\left\{1 - \left(\frac{a\bar{\gamma}_\ell}{1+a\bar{\gamma}_\ell}\right)^b\right\}$ which agrees with the four results given in [1] for BFSK ($a = 1/2$ and $b = 1/2$), BPSK ($a = 1$ and $b = 1/2$), non-coherent BFSK (NC-BFSK) ($a = 1/2$ and $b = 1$) and differentially encoded BPSK (BDPSK) ($a = 1$ and $b = 1$). In addition to these expected ABEP consequences of (19), the AC of the GNM fading channels can be also easily obtained by setting $b = 1$ and $n = 2$ in (19), yielding

$$P_{AC} = \frac{1}{\Gamma(m_\ell)} H_{2,3}^{3,1}\left[\frac{\beta_\ell}{a\bar{\gamma}_\ell} \left| \begin{array}{c} (0,1),(1,1) \\ (0,1),(0,1),(m_\ell, 1/\xi_\ell) \end{array}\right.\right], \qquad (22)$$

where $a$ denotes the transmitted power. Setting $\xi_\ell = 1$ and then using [7, Eq.(8.3.2/21)], (22) further reduces to the well-known result $P_{AC} = G_{2,3}^{3,1}\left[\frac{m_\ell}{a\bar{\gamma}_\ell} \left| \begin{array}{c} 0,1 \\ 0,0,m_\ell \end{array}\right.\right]/\Gamma(m_\ell)$ which is the AC of the Nakagami-$m$ fading channels [14, Eq. (3)]. Furthermore, using [11, Eq. (2.1.7)] and [7, Eq. (8.4.11.3)], then recalling the relation between the first order $E_n$ integral $E_1(x)$ and exponential integral $Ei(x)$ such as $E_1(x) = -Ei(-x)$, for Rayleigh fading channel ($m_\ell = \xi_\ell = 1$), (22) simplifies to [1, Eq. (15.26)] as expected.

In addition to the review above regarding the unified performance measures in GNM fading channels, let us consider the shadowing identified as a main cause not only for reducing energy but also causing performance loss and instability at the receiver. Specifically, shadowing is the effect that the received signal power, i.e., the mean of instantaneous SNR fluctuates due to objects obstructing the propagation path between transmitter and receiver. Additionally, in the architecture of next-generation wireless communication systems, a spectrum well above $60$ GHz frequency will be used, causing greater susceptibility to shadowing. In this context, the modeling of shadowing plays an important role in the context of designing systems and evaluating the corresponding performance. The most well-known shadowing distribution in the literature is the Lognormal distribution [1, and the references therein].

**Example 2** (Unified Performance Measure in Lognormal Fading Channels)**.** The PDF of lognormal distribution can be well approximated in the form of (16) as shown in the second row of Table IV

such that

$$p_{\gamma_\ell}(\gamma) = \frac{1}{\sqrt{\pi}} \sum_{k=1}^{K} \frac{w_k}{\omega_k} \mathrm{H}_{0,0}^{0,0}\left[\frac{\gamma}{\omega_k} \bigg| \begin{array}{c} - \\ - \end{array}\right] \qquad (23)$$

where $\mu_\ell(\text{dB})$ and $\sigma_\ell(\text{dB})$ are the mean and the standard deviation of $\gamma_\ell$, and where $\omega_k$ is defined as $\omega_k = 10^{(\sqrt{2}\sigma_\ell u_k + \mu_\ell)/10}$ such that for $k \in \{1, 2, \ldots, K\}$, $\{w_k\}$ and $\{u_k\}$ are the weight factors and the zeros (abscissas) of the $K$-order Hermite polynomial [6, Table 25.10]. Mapping the parameters and coefficients between (23) and (14), one can readily give the unified performance measure over lognormal fading channels by means of utilizing (16) and then using [7, Eq.(8.3.2/21)] as

$$P_{AUP} = 1 - \frac{n}{2}\left\{1 - \frac{(-1)^n}{\sqrt{\pi}\,\Gamma(b)} \sum_{n=1}^{K} w_k\, \mathrm{G}_{2,n}^{n,1}\left[\frac{1}{a\,\omega_k} \bigg| \begin{array}{c} 1-b,\, 1 \\ \Lambda_{(0)}^{(n)} \end{array}\right]\right\}. \qquad (24)$$

As expected, setting $n=1$ in (24) results in $P_{ABEP} = \frac{1}{2\sqrt{\pi}} \sum_{n=1}^{K} w_k \Gamma(b, a\omega_k)/\Gamma(b)$. Moreover, setting $b=1$ and $n=2$, (24) simplifies into $P_{ABEP} = \frac{1}{2\sqrt{\pi}} \sum_{n=1}^{K} w_k \log(1 + a\omega_k)$ as expected.

In addition to the context of shadowing, composite fading channels (include both shadowing and fading) plays an important role in designing and modeling wireless communication systems. To the best of our knowledge, another general composite fading model is the extended generalized-K (EGK) fading. In the following example, unified performance measure $P_{AUP}$ is analyzed over EGK fading channels [15].

**Example 3** (Unified Performance Measure in Extended Generalized-K (EGK) Fading Channels). The distribution of the instantaneous SNR in the EGK fading channel follows the PDF given by [15, Eq. (3)], [16], that is,

$$p_{\gamma_\ell}(\gamma) = \frac{\xi_\ell}{\Gamma(m_\ell)\Gamma(m_{s\ell})} \left(\frac{\beta_\ell \beta_{s\ell}}{\Omega_{s\ell}}\right)^{\xi_\ell m_\ell} \gamma^{\xi_\ell m_\ell - 1} \Gamma\left(m_{s\ell} - m_\ell \frac{\xi_\ell}{\xi_{s\ell}}, 0, \left(\frac{\beta_\ell \beta_{s\ell}}{\Omega_{s\ell}}\right)^{\xi_\ell} \gamma^{\xi_\ell}, \frac{\xi_\ell}{\xi_{s\ell}}\right) \qquad (25)$$

defined over $0 \leq \gamma < \infty$, where the parameters $m_\ell$ ($0.5 \leq m_\ell < \infty$) and $\xi_\ell$ ($0 \leq \xi_\ell < \infty$) represent the fading figure (diversity severity / order) and the fading shaping factor, respectively, while $m_{s\ell}$ ($0.5 \leq m_{s\ell} < \infty$) and $\xi_{s\ell}$ ($0 \leq \xi_{s\ell} < \infty$) represent the shadowing severity and the shadowing shaping factor (inhomogeneity), respectively. In addition, the parameters $\beta_\ell$ and $\beta_{s\ell}$ are defined as $\beta_\ell = \Gamma(m_\ell + 1/\xi_\ell)/\Gamma(m_\ell)$ and $\beta_{s\ell} = \Gamma(m_{s\ell} + 1/\xi_{s\ell})/\Gamma(m_{s\ell})$, respectively. In addition, $\Gamma(\cdot, \cdot, \cdot, \cdot)$ is the extended incomplete Gamma function defined as $\Gamma(\alpha, x, b, \beta) = \int_x^\infty r^{\alpha - 1} \exp\left(-r - br^{-\beta}\right) dr$, where $\alpha, \beta, b \in \mathbb{C}$ and $x \in \mathbb{R}^+$ [17, Eq. (6.2)]. The Fox's H representation of (25) is given by at the second row

of Table V, that is

$$p_{\gamma_\ell}(\gamma) = \frac{\beta_\ell \beta_{s\ell}}{\Gamma(m_\ell)\Gamma(m_{s\ell})\bar{\gamma}_{s\ell}} \mathrm{H}_{0,2}^{2,0}\left[\frac{\beta_\ell \beta_{s\ell}}{\bar{\gamma}_{s\ell}}\gamma \,\middle|\, \overline{\phantom{xx}} \atop (m_\ell - 1/\xi_\ell,\, 1/\xi_\ell),\, (m_{s\ell} - 1/\xi_{s\ell},\, 1/\xi_{s\ell})\right]. \qquad (26)$$

Mapping the parameters and coefficients of (26) to those of the hyper-Fox's H fading model defined in Definition 1 with $K = 1$, the unified performance measure $P_{AUP}$ can be readily expressed as

$$P_{AUP} = 1 - \frac{n}{2}\left\{1 - \frac{(-1)^n}{\Gamma(b)\Gamma(m_\ell)\Gamma(m_{s\ell})} \mathrm{H}_{2,n+2}^{n+2,1}\left[\frac{\beta_\ell \beta_{s\ell}}{a\,\bar{\gamma}_{s\ell}} \,\middle|\, {(1-b,1),(1,1) \atop \Lambda_{((0,1))}^{(n)},\, (m_\ell,\tfrac{1}{\xi_\ell}),\, (m_{s\ell},\tfrac{1}{\xi_{s\ell}})}\right]\right\}. \qquad (27)$$

Note that (27) reduces to the ABEP in EGK fading channels [16, Eq. (34)] by means of setting $n = 1$ and using the same steps in the derivation of (21). Moreover, for $n = 2$ and $b = 1$, (27) simplifies into the AC in EGK fading channels [16, Eq. (42)] as expected.

In order to check the analytical accuracy and correctness, some numerical and simulation results regarding the ABEP and AC performances of single link reception over fading channels are depicted in Fig. 1, Fig. 2 and Fig. 3, and they show that these numerical and simulation results are in perfect agreement.

### B. Multiple Link Reception

We consider an $L$-branch maximal ratio combiner (MRC) diversity system employing binary modulation and operating in a slow non-selective mutual independent and not-necessarily identically distributed generalized fading environment corrupted by AWGN noise. The instantaneous SNR $\gamma_{end}$ at the output of the MRC receiver is considered as the sum of the instantaneous SNRs of the branches, that is,

$$\gamma_{end} = \sum_{\ell=1}^{L} \gamma_\ell, \qquad (28)$$

where $L$ denotes the number of branches, and where for $\ell \in \{1, 2, 3, \ldots, L\}$, $\gamma_\ell$ is the instantaneous SNR the $\ell$th branch is subjected to. The AUP expression $P_{AUP}$ of the $L$-branch MRC combiner can be obtained by averaging the (instantaneous) unified performance measure $P_{UP}(\gamma_{end})$ given by Corollary 4 over the PDF of $\gamma_{end} = \sum_{\ell=1}^{L} \gamma_\ell$ as shown in (12). Due to several reasons (e.g., insufficient antenna spacing or coupling among RF layers), correlation may exist among diversity branches of the $L$-branch MRC. With that, the AC using (12) involves an $L$-fold integral given by

$$P_{AUP} = 1 - \frac{n}{2}\left\{1 - \frac{(-1)^n}{\Gamma(b)}\underbrace{\int_0^\infty \int_0^\infty \cdots \int_0^\infty}_{L\text{-fold}} G_{n,2}^{1,n}\left[a\sum_{\ell=1}^L r_\ell \left| \begin{array}{c} \Lambda_1^{(n)} \\ b, 0 \end{array}\right.\right] \times \right.$$
$$\left. p_{\gamma_1,\gamma_2,\ldots,\gamma_L}(r_1, r_2, \ldots, r_L) dr_1 dr_2 \ldots dr_L \right\}, \quad (29)$$

where $p_{\gamma_1,\gamma_2,\ldots,\gamma_L}(r_1, r_2, \ldots, r_L)$ is the joint multivariate PDF of the instantaneous SNRs $\{\gamma_\ell\}_{\ell=1}^L$. In (29), the $L$-fold integration is tedious and cannot be partioned into the product of one dimensional integrals even if the instantaneous SNRs $\{\gamma_\ell\}_{\ell=1}^L$ are assumed mutually independent. Additionally, it is clear that the numerical evaluation of (29) is complex requiring a long time to compute the desired result as the number of branches $L$ increases. Fortunately, after performing some algebraic manipulations, the AUP expression can be readily obtained in terms of a single integral expression using an MGF-based approach [1] as presented in the following theorem.

**Theorem 3** (Average Unified Expression of the $L$-branch Diversity Combiners over Correlated Not-Necessarily Identically Distributed Fading Channels). *The exact AUP of L-branch diversity combiner over mutually not-necessarily independent nor identically distributed fading channels is given by*

$$P_{AUP} = 1 - \frac{n}{2}\left\{1 + \frac{(-1)^n}{\Gamma(b)}\int_0^\infty G_{n+1,2}^{1,n}\left[\frac{a}{s}\left|\begin{array}{c}\Lambda_1^{(n)}, 1 \\ b, 0\end{array}\right.\right]\left[\frac{\partial}{\partial s}\mathcal{M}_{\gamma_1,\gamma_2,\ldots,\gamma_L}(s)\right]ds\right\}, \quad (30)$$

*where the parameters $a \in \mathbb{R}^+$, $b \in (0,1]$, $n \in \{1,2\}$ are selected according to desired performance measure, and where $\mathcal{M}_{\gamma_1,\gamma_2,\ldots,\gamma_L}(s) \equiv \mathbb{E}[\exp(-s\sum_\ell \gamma_\ell)]$ is the joint MGF of the correlated instantaneous SNRs $\gamma_1, \gamma_2, \ldots, \gamma_L$ of the branches.*

*Proof:* See Appendix C. ∎

It is worth accentuating that, in order to find the AUP expression of the diversity combiner, the MGF-based technique proposed in Theorem 3 eliminates the necessity of finding the PDF of the instantaneous SNR $\gamma_{end} = \sum_{\ell=1}^L \gamma_\ell$ through the inverse Laplace transform (ILT) of the joint MGF $\mathcal{M}_{\gamma_1,\gamma_2,\ldots,\gamma_L}(s) \equiv \mathbb{E}[\exp(-s\sum_\ell \gamma_\ell)]$. Shortly, Theorem 3 suggests that one can readily obtain the AUP expression using the joint MGF $\mathcal{M}_{\gamma_1,\gamma_2,\ldots,\gamma_L}(s)$. Additionally, the integral in (30) can be accurately estimated by employing the Gauss-Chebyshev Quadrature (GCQ) formula [6, Eq.(25.4.39)], i.e.,

$$P_{AUP} = 1 - \frac{n}{2}\left\{1 + \frac{(-1)^n}{\Gamma(b)}\sum_{n=1}^N w_n G_{n+1,2}^{1,n}\left[\frac{a}{s_n}\left|\begin{array}{c}\Lambda_1^{(n)}, 1 \\ b, 0\end{array}\right.\right]\left\{\frac{\partial}{\partial s}\mathcal{M}_{\gamma_1,\gamma_2,\ldots,\gamma_L}(s)\bigg|_{s=s_n}\right\}\right\}, \quad (31)$$

which converges rapidly and steadily, requiring only few terms for an accurate result. In (31), the coefficients $w_n$ and $s_n$ are defined as

$$w_n = \frac{\pi^2 \sin\left(\frac{2n-1}{2N}\pi\right)}{4N\cos^2\left(\frac{\pi}{4}\cos\left(\frac{2n-1}{2N}\pi\right) + \frac{\pi}{4}\right)} \quad \text{and} \quad s_n = \tan\left(\frac{\pi}{4}\cos\left(\frac{2n-1}{2N}\pi\right) + \frac{\pi}{4}\right), \quad (32)$$

respectively, where the truncation index $N$ could be chosen more than $N = 30$ to obtain a high level of accuracy. Despite the fact that the novel technique represented by Theorem 3 is easy to use, and referring to both (30) and (31), let us consider its special cases in order to check its analytical simplicity and accuracy.

**Special Case 1** (Average Bit Error Probabilities of $L$-Branch Diversity Combiner). As mentioned before, the ABEP $P_{ABEP} = \frac{1}{2}\mathbb{E}[\Gamma(b, a\gamma_{end})/\Gamma(b)]$ of the $L$-branch diversity combiner can be readily obtained through setting $n = 1$ in (30), such that $P_{UP}|_{n=1}$ results in

$$P_{ABEP} = \frac{1}{2} + \frac{1}{2\Gamma(b)} \int_0^\infty G_{2,2}^{1,1}\left[\frac{a}{s} \middle| \begin{array}{c} 1, 1 \\ b, 0 \end{array}\right] \left[\frac{\partial}{\partial s}\mathcal{M}_{\gamma_1,\gamma_2,\ldots,\gamma_L}(s)\right] ds. \quad (33)$$

To the best of the authors' knowledge, the result given by (34) is a new result which readily simplifies into the performance of BFSK ($a = 1/2$ and $b = 1/2$), BPSK ($a = 1$ and $b = 1/2$), non-coherent BFSK ($a = 1/2$ and $b = 1$) and BDPSK ($a = 1$ and $b = 1$).

**Special Case 2** (Average Capacity of $L$-Branch Diversity Combiner). The AC $P_{AC} = \mathbb{E}[\log(1 + a\gamma_{end})]$ of the $L$-branch diversity combiner can be readily obtained through setting $b = 1$ and $n = 2$ in (30), such that $P_{UP}|_{n=2,b=1}$ results in

$$P_{AC} = -\int_0^\infty G_{3,2}^{1,2}\left[\frac{a}{s} \middle| \begin{array}{c} 1, 1, 1 \\ 1, 0 \end{array}\right] \left[\frac{\partial}{\partial s}\mathcal{M}_{\gamma_1,\gamma_2,\ldots,\gamma_L}(s)\right] ds, \quad (34)$$

where it may be useful to notice that $G_{3,2}^{1,2}\left[\frac{a}{s}\middle|\begin{array}{c}1,1,1\\1,0\end{array}\right] = G_{2,1}^{0,2}\left[\frac{a}{s}\middle|\begin{array}{c}1,1\\0\end{array}\right]$ by means of employing [7, Eq. (8.2.2/9)]. Then, using both [7, Eq. (8.2.2/14)] and $G_{2,1}^{0,2}\left[\frac{a}{s}\middle|\begin{array}{c}1,1\\0\end{array}\right] = -\mathrm{Ei}\left(-\frac{s}{a}\right)$ [7, Eq. (8.4.11/1)], where $\mathrm{Ei}(\cdot)$ is the exponential integral function [8, Eq. (8.211/1)], (34) simplifies to

$$P_{AC} = \int_0^\infty \mathrm{Ei}(-s/a)\left[\frac{\partial}{\partial s}\mathcal{M}_{\gamma_1,\gamma_2,\ldots,\gamma_L}(s)\right] ds, \quad (35)$$

which is a well-known result given by Di Renzo *et al.* in [3, Eq. (7)].

Note that the spatial correlation between all fading amplitudes can be determined from the physical parameters of the model, which includes antenna spacing, antenna arrangement, angle spread, and angle of arrival. In the case of there does not exist any correlation between all fading amplitudes $\gamma_1, \gamma_2, \ldots, \gamma_L$ for the branches of the diversity combiner, the AUP expression is given in the following corollary.

**Corollary 5** (Average Unified Expression of the $L$-branch Diversity Combiners over Mutually Independent Non-Identically Distributed Fading Channels). *The exact AUP expression of $L$-branch MRC over mutually independent and non-identically distributed fading channels is given by*

$$P_{AUP} = 1 - \frac{n}{2}\left\{1 + \frac{(-1)^n}{\Gamma(b)}\int_0^\infty G_{n+1,2}^{1,n}\left[\frac{a}{s}\,\middle|\,\begin{array}{c}\Lambda_1^{(n)}, 1 \\ b, 0\end{array}\right]\sum_{\ell=1}^{L}\left[\frac{\partial}{\partial s}\mathcal{M}_{\gamma_\ell}(s)\right]\prod_{\substack{k=1 \\ k\neq\ell}}^{L}\mathcal{M}_{\gamma_k}(s)\,ds\right\}, \qquad (36)$$

*where, for $\ell \in \{1, 2, \ldots, L\}$, $\mathcal{M}_{\gamma_\ell}(s) \equiv \mathbb{E}[\exp(-s\gamma_\ell)]$ is the MGF of the instantaneous SNR $\gamma_\ell$ that the $\ell$th branch is subjected to.*

*Proof:* When there is no correlation between all instantaneous SNRs $\gamma_1, \gamma_2, \ldots, \gamma_L$ of the branches, one can readily write $\mathcal{M}_{\gamma_1,\gamma_2,\ldots,\gamma_L}(s) = \prod_{\ell=1}^{L}\mathcal{M}_{\gamma_\ell}(s)$ whose derivation with respect to $s$ is

$$\frac{\partial}{\partial s}\mathcal{M}_{\gamma_1,\gamma_2,\ldots,\gamma_L}(s) = \sum_{\ell=1}^{L}\left[\frac{\partial}{\partial s}\mathcal{M}_{\gamma_\ell}(s)\right]\prod_{\substack{k=1 \\ k\neq\ell}}^{L}\mathcal{M}_{\gamma_k}(s). \qquad (37)$$

Finally, substituting (37) into (30) results in (36), which proves Corollary 5. ∎

As mentioned before and nicely shown in Tables II, III, IV and V, the PDF of several non-negative distributions can be compactly expressed or accurately approximated in the form of (14). Using either [12, Eq. (2.12)] or [18, Eq. (2.10)], the MGF of the hyper-Fox's H fading channel is given by

$$\mathcal{M}_{\gamma_\ell}(s) = \sum_{n_\ell=1}^{K_\ell}\frac{\eta_{n_\ell}}{c_{n_\ell}}\,\mathrm{H}_{P_{n_\ell}+1,Q_{n_\ell}}^{M_{n_\ell},N_{n_\ell}+1}\left[\frac{c_{n_\ell}}{s}\,\middle|\,\begin{array}{l}(1,1),(\vartheta_{n_\ell j}+\theta_{n_\ell j},\theta_{n_\ell j})_{j=1}^{P_{n_\ell}} \\ (\varphi_{n_\ell j}+\phi_{n_\ell j},\phi_{n_\ell j})_{j=1}^{Q_{n_\ell}}\end{array}\right] \qquad (38)$$

with the convergence region $\Re\{s\} \geq 0$. Then, with the aid of [7, Eq. (8.3.2/15)], the derivative of (38) can be readily derived as

$$\frac{\partial}{\partial s}\mathcal{M}_{\gamma_\ell}(s) = -\sum_{n_\ell=1}^{K_\ell}\frac{\eta_{n_\ell}}{c_{n_\ell}s}\,\mathrm{H}_{P_{n_\ell}+1,Q_{n_\ell}}^{M_{n_\ell},N_{n_\ell}+1}\left[\frac{c_{n_\ell}}{s}\,\middle|\,\begin{array}{l}(0,1),(\vartheta_{n_\ell j}+\theta_{n_\ell j},\theta_{n_\ell j})_{j=1}^{P_{n_\ell}} \\ (\varphi_{n_\ell j}+\phi_{n_\ell j},\phi_{n_\ell j})_{j=1}^{Q_{n_\ell}}\end{array}\right] \qquad (39)$$

Finally, substituting both (38) and (39) into (36), the AUP expression of the $L$-branch diversity combiner performing over hyper-Fox's H fading channel can be readily obtained in the form of Corollary 5 as

$$P_{AUP} = 1 - \frac{n}{2}\left\{1 - \frac{(-1)^n}{\Gamma(b)}\int_0^\infty G_{n+1,2}^{1,n}\left[\frac{a}{s}\left|\begin{array}{c}\Lambda_1^{(n)},1\\b,0\end{array}\right.\right]\times\right.$$
$$\sum_{\ell=1}^L\sum_{n_\ell=1}^{K_\ell}\frac{\eta_{n_\ell}}{c_{n_\ell}s}H_{P_{n_\ell}+1,Q_{n_\ell}}^{M_{n_\ell},N_{n_\ell}+1}\left[\frac{c_{n_\ell}}{s}\left|\begin{array}{c}(0,1),(\vartheta_{n_\ell j}+\theta_{n_\ell j},\theta_{n_\ell j})_{j=1}^{P_{n_\ell}}\\(\varphi_{n_\ell j}+\phi_{n_\ell j},\phi_{n_\ell j})_{j=1}^{Q_{n_\ell}}\end{array}\right.\right]\times$$
$$\left.\prod_{\substack{k=1\\k\neq\ell}}^L\sum_{n_k=1}^{K_k}\frac{\eta_{n_k}}{c_{n_k}}H_{P_{n_k}+1,Q_{n_k}}^{M_{n_k},N_{n_k}+1}\left[\frac{c_{n_k}}{s}\left|\begin{array}{c}(1,1),(\vartheta_{n_k j}+\theta_{n_k j},\theta_{n_k j})_{j=1}^{P_{n_k}}\\(\varphi_{n_k j}+\phi_{n_k j},\phi_{n_k j})_{j=1}^{Q_{n_k}}\end{array}\right.\right]ds\right\}. \quad (40)$$

Let us consider some special cases of the hyper-Fox's H fading model, i.e., the special cases of (40). Note that the MGFs of some commonly used fading distributions (such as one-sided Gaussian, exponential, Gamma, Weibull, hyper-Gamma, Nakagami-$q$ (Hoyt), Nakagami-$n$ (Rice), Maxwell, lognormal, K-distribution, generalized-K, generalized Gamma, extended generalized Gamma and Fox's H) and their derivatives are given in details in Tables II,III,IV and V. Using the MGF of generalized Gamma distribution and its derivative given in the first row in Table V, and then substituting them into (40), the AUP expression of the $L$-branch combiner over GNM fading channels is given by

$$P_{AUP} = 1 - \frac{n}{2}\left\{1 - \frac{(-1)^n}{\Gamma(b)}\left[\prod_{k=1}^L\frac{1}{\Gamma(m_k)}\right]\int_0^\infty \frac{1}{s}G_{n+1,2}^{1,n}\left[\frac{a}{s}\left|\begin{array}{c}\Lambda_1^{(n)},1\\b,0\end{array}\right.\right]\times\right.$$
$$\left.\sum_{\ell=1}^L H_{1,1}^{1,1}\left[\frac{\beta_\ell}{\bar{\gamma}_\ell s}\left|\begin{array}{c}(0,1)\\(m_\ell,\frac{1}{\xi_\ell})\end{array}\right.\right]\prod_{\substack{k=1\\k\neq\ell}}^L H_{1,1}^{1,1}\left[\frac{\beta_k}{\bar{\gamma}_k s}\left|\begin{array}{c}(1,1)\\(m_k,\frac{1}{\xi_k})\end{array}\right.\right]ds\right\}. \quad (41)$$

Note that in the special case of single link, i.e., $L = 1$, (41) not surprisingly simplifies into (19) by means of some algebraic manipulations using [7, Eq.(2.25.1/1)] and [7, Eq.(8.3.2/21)] together. Some algebraic manipulations substituting the shaping parameters $\xi_1 = \xi_2 = \cdots = \xi_N = 1$ into (41) and then utilizing [7, Eqs. (8.3.2/7), (8.3.2/21) and then (8.4.2/5)] result in the unified expression of the $L$-branch diversity combiner performing over mutually independent and not necessarily identically distributed Nakagami-$m$ fading channels, that is,

$$P_{AUP} = 1 - \frac{n}{2}\left\{1 - \frac{(-1)^n}{\Gamma(b)}\int_0^\infty G_{n+1,2}^{1,n}\left[\frac{a}{s}\left|\begin{array}{c}\Lambda_1^{(n)},1\\b,0\end{array}\right.\right]\frac{\sum_{\ell=1}^L\frac{\bar{\gamma}_\ell}{1+\frac{\bar{\gamma}_\ell s}{m_\ell}}}{\prod_{\ell=1}^L\left(1+\frac{\bar{\gamma}_\ell s}{m_\ell}\right)^{m_\ell}}ds\right\}. \quad (42)$$

For identically distributed Nakagami-$m$ fading channels (i.e., $m_1 = m_2 = \ldots = m_L = m$ and $\bar{\gamma}_1 = \bar{\gamma}_2 = \ldots = \bar{\gamma}_L = \bar{\gamma}$), apparently as a result of both employing $\Gamma(mL+1)\left(1+\frac{\bar{\gamma}s}{m}\right)^{-mL-1} = G_{1,1}^{1,1}\left[\frac{\bar{\gamma}s}{m}\left|\begin{array}{c}-mL\\0\end{array}\right.\right]$

[7, Eq.(8.4.2/5)] and then using the integral equality of two Meijer's G functions [7, Eq.(2.24.1/1)], (42) simplifies to

$$P_{AUP} = 1 - \frac{n}{2}\left\{1 - \frac{(-1)^n}{\Gamma(b)\Gamma(mL)} G_{2,n+1}^{n+1,1}\left[\frac{m}{a\bar{\gamma}} \middle| \begin{matrix} 1-b, 1 \\ \Lambda_0^{(n)}, mL \end{matrix}\right]\right\} \qquad (43)$$

which is the special case of (19) with $\xi_\ell = 1, m_\ell = m, \bar{\gamma}_\ell = \bar{\gamma}$ and $L = 1$ as it is expected. In addition, for $n = 1$, following the same steps in the derivation of (21) from (20), the unified expression given by (42) readily reduces to the ABEP of the $L$-branch diversity combiner over identical Nakagami-$m$ fading channels can be readily obtained as

$$P_{ABEP} = \frac{1}{2\Gamma(b)\Gamma(mL)} G_{2,2}^{1,2}\left[\frac{m}{a\bar{\gamma}} \middle| \begin{matrix} 1, 1-b \\ 0, mL \end{matrix}\right]. \qquad (44)$$

For $n = 2$ and $b = 1$, (43) simplifies to the well-known result [4, Eq. (33)], that is,

$$P_{AC} = \frac{1}{\Gamma(mL)} G_{2,3}^{3,1}\left[\frac{m}{a\bar{\gamma}} \middle| \begin{matrix} 0, 1 \\ 0, 0, mL \end{matrix}\right], \qquad (45)$$

where $a$ denotes the transmitted power.

As an illustration of the mathematical formalism presented above, some numerical and simulation results regarding the ABEP and AC performance of multiple link reception over fading channels are depicted in Fig. 4, Fig. 5 and Fig. 6, and these figures show that these analytical and simulation results are in perfect agreement.

## IV. CONCLUSION

In this paper, we presented a unified performance expression combining the ABEP and AC of wireless communication systems over generalized fading channels. More precisely, this paper introduces an MGF-based unified expression for the ABEP and AC of single and multiple link communication with an $L$-branch MRC combining. In addition, the hyper-Fox's H fading model is proposed as a unified fading distribution for a majority of the well-known generalized fading models in order to provide more general and more generic results which can be readily simplified to some published results for some well-known fading distributions. We explicitly offer a generic unified performance expression that can be easily calculated and that is applicable to a wide variety of fading scenarios. Finally, as an illustration of the mathematical formalism, some simulations have been carried out for different scenarios of fading environment, and numerical and simulation results were shown to be in perfect agreement.

## ACKNOWLEDGMENTS

This work is supported by King Abdullah University of Science and Technology (KAUST).

## APPENDIX A

### UNIFIED BEP EXPRESSION USING THE INCOMPLETE BETA FUNCTION

Note that, substituting the lower incomplete gamma $\gamma(a, z) = \int_0^z e^{-u} u^{a-1} du$ [6, Eq. (6.5.2)] and using the relation between the lower and upper incomplete Gamma functions (i.e., $\gamma(a, z) = \Gamma(a) - \Gamma(a, z)$) [6, Eq. (6.5.2)], the conditional bit error probability $P_{BER}(\gamma_{end})$ can be written as

$$P_{BEP}(\gamma_{end}) = \frac{1}{2} - \frac{\gamma(b, a\gamma_{end})}{2\Gamma(b)}. \tag{A.1}$$

Substituting an alternative representation of the lower incomplete gamma function $\gamma(a, z) = b^{-1}(az)^b {}_1F_1[a; a+1; -z]$ [7, Eq. (7.11.3/1)] in (A.1) and then using the well-known limit representation ${}_1F_1[-z; a; a+1] = \lim_{d \to \infty} {}_2F_1[d, a; a+1; -z/d]$ [7, Eq. (7.2.2/13)], we can write

$$P_{BEP}(\gamma_{end}) = \frac{1}{2} - \frac{(a\gamma_{end})^b}{2\Gamma(b+1)} \lim_{d \to \infty} {}_2F_1\left[d, b; b+1; -\frac{a}{d}\gamma_{end}\right], \tag{A.2}$$

where ${}_1F_1[\cdot; \cdot; \cdot]$ and ${}_2F_1[\cdot; \cdot, \cdot; \cdot]$ are the Kummer confluent and Gauss hypergeometric functions, respectively. Finally, substituting the Gauss hypergeometric representation of the incomplete beta function, i.e., ${}_2F_1[z; d, b; b+1] = bz^{-b} B(z; b, 1-d)$ [7, Eq. (7.3.1/28)] into (A.2) results in (2), which proves Theorem 1.

## APPENDIX B

### CAPACITY EXPRESSION USING THE INCOMPLETE BETA FUNCTION

Using [7, Eq. (7.3.3/7)], i.e.,

$$_2F_1\left[\frac{m}{n}, 1; \frac{m}{n} + 1; -z\right] = -\frac{m}{n} z^{-m/n} \sum_{k=0}^{n-1} e^{-i(2k+1)\frac{m}{n}\pi} \log\left(1 - z^{m/n} e^{i(2k+1)\frac{m}{n}\pi}\right), \tag{B.1}$$

one can readily show by setting $m = n = 1$ in (B.1) that the conditional capacity $P_C(\gamma_{end})$ can also be represented as

$$P_C(\gamma_{end}) = \gamma_{end} \, {}_2F_1[1, 1; 2; -\gamma_{end}]. \tag{B.2}$$

Substituting the Gauss hypergeometric representation of the incomplete beta function, i.e., ${}_2F_1[z; a, b; b+1] = bz^{-b} B(z; b, 1-a)$ [7, Eq. (7.3.1/28)] into (B.2) results in (4), which proves Theorem 2.

# APPENDIX C

## AVERAGE UNIFIED EXPRESSION OF THE $L$-BRANCH DIVERSITY COMBINERS OVER CORRELATED NOT-NECESSARILY IDENTICALLY DISTRIBUTED FADING CHANNELS

We utilize the MacRobert's E function representation of the UP expression (i.e., see Corollary 3) instead of the other representations. More specifically, using the well-known the integral formula of the MacRobert's E function [19, Eq. (6)], [8, Eq. (7.814/2)], given by

$$\int_0^\infty \exp(-u) u^{\alpha_{p+1}-1} E\left[\{\alpha_i\}_{i=1}^p; \{\rho_j\}_{j=1}^q; \frac{z}{u}\right] du = E\left[\{\alpha_i\}_{i=1}^{p+1}; \{\rho_j\}_{j=1}^q; z\right] \quad \text{(C.1)}$$

where $\Re\{\alpha_{p+1}\} > 0$ and $z \in \mathbb{R}^+$, and performing some algebraic manipulations, one can readily obtain an alternative representation of (C.1) as

$$(a\gamma_{end})^b E\left[\Lambda_b^{(n)}; b+1; \frac{1}{a\gamma_{end}}\right] = \int_0^\infty u^{b-1} e^{-\frac{u}{a\gamma_{end}}} E\left[\Lambda_b^{n-1}; b+1; \frac{1}{u}\right] du. \quad \text{(C.2)}$$

Accordingly, substituting (C.2) and the total instantaneous SNR $\gamma_{end} = \sum_{\ell=1}^L \gamma_\ell$ into (8), it is straightforward to show that the AUP expression $P_{AUP} = \mathbb{E}[P_{UP}(\gamma_{end})]$ for an $L$-branch diversity combiner can be explicitly given by

$$P_{AUP} = 1 - \frac{n}{2}\left\{1 - \frac{(-1)^n}{\Gamma(b)} \int_0^\infty u^{b-1} \mathbb{E}\left[\exp\left(-\frac{u}{a\sum_{\ell=1}^L \gamma_\ell}\right)\right] E\left[\Lambda_b^{(n-1)}; b+1; \frac{1}{u}\right] du\right\}. \quad \text{(C.3)}$$

Note that, $\mathbb{E}\left[\exp(-s/\sum_{\ell=1}^L \gamma_\ell)\right]$ with $s = u/a$ can be considered as the MGF of the reciprocal distribution of $\gamma_{end} = \sum_{\ell=1}^L \gamma_\ell$. In order to proceed further, we present the following lemma.

**Lemma 1** (MGF of Reciprocal Distribution [20, Theorem 1]). *Given any nonnegative continuous RV $\mathcal{R}$ distributed over $(0, \infty)$ with the PDF $p_\mathcal{R}(r)$ and the MGF $\mathcal{M}_\mathcal{R}(s)$ for $\Re\{s\} \geq 0$. Then, the MGF of its reciprocal distribution $\widetilde{\mathcal{R}}$ (i.e., $\widetilde{\mathcal{R}} \equiv 1/\mathcal{R}$) is given by*

$$\mathcal{M}_{\widetilde{\mathcal{R}}}(s) = -\int_0^\infty J_0\left(2\sqrt{su}\right)\left[\frac{\partial}{\partial u}\mathcal{M}_\mathcal{R}(u)\right] du \quad \text{(C.4)}$$

*with the convergence region $\Re\{s\} \geq 0$, where $J_0(\cdot)$ is the zeroth-order Bessel function of the first kind defined in [6, Eq. (6.19.7)].*

Accordingly, applying Lemma 1 on the expectation part $\mathbb{E}\left[\exp(-(u/a)/\sum_{\ell=1}^{L}\gamma_\ell)\right]$ of (C.3) results as

$$P_{AUP} = 1 - \frac{n}{2}\left\{1 + \frac{(-1)^n}{\Gamma(b)}\int_0^\infty\int_0^\infty u^{b-1}J_0\left(2\sqrt{\frac{up}{a}}\right)\times \right.$$
$$\left. E\left[\Lambda_b^{(n-1)};b+1;\frac{1}{u}\right]du\left[\frac{\partial}{\partial p}\exp\left(-p\sum_{\ell=1}^{L}\gamma_\ell\right)\right]dp\right\}. \quad \text{(C.5)}$$

Note that using the Hankel transform of the MacRobert's E function [8, Eq. (7.823/1)], and then applying [21, Eq. (1)], the inner integral of (C.5) can be easily obtained in closed-form as

$$\int_0^\infty u^{b-1}J_0\left(2\sqrt{\frac{up}{a}}\right)E\left[\Lambda_b^{(n-1)};b+1;\frac{1}{u}\right]du = G_{n+1,2}^{1,n}\left[\frac{a}{p}\left|\begin{array}{c}\Lambda_1^{(n)},1\\b,0\end{array}\right.\right]. \quad \text{(C.6)}$$

Finally, replacing the inner integral in (C.5) with the Hankel transform of MacRobert's E function given by (C.6) and using the definition of joint MGF (i.e., $\mathcal{M}_{\gamma_1,\gamma_2,\ldots,\gamma_L}(p) \equiv \mathbb{E}[\exp(-p\sum_\ell \gamma_\ell)]$), (C.5) simplifies to (30), which proves Theorem 3.

TABLE I
PARAMETERS $a$, $b$, $n$ AND $d$ FOR THE BEP AND CAPACITY PERFORMANCE MEASURES

| Performance Measure | $a$ | $b$ | $n$ | $d$ |
|---|---|---|---|---|
| BEP of orthogonal coherent BFSK [1, Eq. (8.43)], $P_{BEP}(\gamma_{end}) = Q\left(\sqrt{\gamma_{end}}\right)$, where $Q(\cdot)$ is the Gaussian Q-function having a one-to-one mapping with the complementary error function $\text{erfc}(\cdot)$, i.e., $Q(z) = \frac{1}{2}\text{erfc}(z/\sqrt{2})$. | $\frac{1}{2}$ | $\frac{1}{2}$ | 1 | $\infty$ |
| BEP of orthogonal noncoherent BFSK [1, Eq. (8.69)], $P_{BEP}(\gamma_{end}) = \frac{1}{2}\exp\left(-\frac{\gamma_{end}}{2}\right)$. | $\frac{1}{2}$ | 1 | 1 | $\infty$ |
| BEP of antipodal coherent BPSK [1, Eq. (8.19)], $P_{BEP}(\gamma_{end}) = Q\left(\sqrt{2\gamma_{end}}\right)$. | 1 | $\frac{1}{2}$ | 1 | $\infty$ |
| BEP of antipodal differentially coherent BPSK (DPSK) [1, Eq. (8.85)], $P_{BEP}(\gamma_{end}) = \frac{1}{2}\exp\left(-\gamma_{end}\right)$. | 1 | 1 | 1 | $\infty$ |
| BEP of correlated coherent binary signaling [1, Chapter 8, footnote 6], $P_{BEP}(\gamma_{end}) = Q\left(\sqrt{2a\gamma_{end}}\right)$. | $a \in [0,1]$ | $\frac{1}{2}$ | 1 | $\infty$ |
| Shannon Capacity [1, Eq. (15.22)], $P_C(\gamma_{end}) = \log(1 + a\gamma_{end})$, where $a$ is the transmitted power and $\log(\cdot)$ is the natural logarithm (i.e., the logarithm to the base $e$) [6, Eq. (4.1.1)]. | $a \in \mathbb{R}^+$ | 1 | 2 | 1 |

TABLE II
MGFs OF SOME WELL-KNOWN FADING CHANNEL MODELS

| **Instantaneous SNR Distribution, i.e., $p_{\gamma_\ell}(\gamma)$** | **MGF $\mathcal{M}_{\gamma_\ell}(s)$ and its derivative $\frac{\partial}{\partial s}\mathcal{M}_{\gamma_\ell}(s)$** |
|---|---|
| **One-Sided Gaussian** [1, Sec. 2.2.1.4] $$p_{\gamma_\ell}(\gamma) = \sqrt{\frac{2}{\pi\bar{\gamma}_\ell}} \exp\left(-\frac{\gamma^2}{2\bar{\gamma}_\ell}\right),$$ $$= \frac{1}{2\sqrt{\pi}\bar{\gamma}_\ell} \mathrm{H}_{0,1}^{1,0}\left[\frac{\gamma}{2\bar{\gamma}_\ell} \middle| \begin{array}{c} \text{---} \\ (-\frac{1}{2},1) \end{array}\right],$$ where $\bar{\gamma}_\ell$ is the average power (i.e., $\bar{\gamma}_\ell \geq 0$). | $$\mathcal{M}_{\gamma_\ell}(s) = \frac{1}{\sqrt{\pi}} \mathrm{H}_{1,1}^{1,1}\left[\frac{1}{2\bar{\gamma}_\ell s} \middle| \begin{array}{c} (1,1) \\ (\frac{1}{2},1) \end{array}\right] = \frac{1}{\sqrt{\pi}} \mathrm{G}_{1,1}^{1,1}\left[\frac{1}{2\bar{\gamma}_\ell s} \middle| \begin{array}{c} 1 \\ \frac{1}{2} \end{array}\right] = \frac{1}{\sqrt{1+2\bar{\gamma}_\ell s}},$$ $$\frac{\partial}{\partial s}\mathcal{M}_{\gamma_\ell}(s) = \frac{2\bar{\gamma}_\ell}{\sqrt{\pi}} \mathrm{H}_{2,2}^{2,1}\left[\frac{1}{2\bar{\gamma}_\ell s} \middle| \begin{array}{c} (2,1),(1,1) \\ (\frac{3}{2},1),(2,1) \end{array}\right] = \frac{2\bar{\gamma}_\ell}{\sqrt{\pi}} \mathrm{G}_{2,2}^{2,1}\left[\frac{1}{2\bar{\gamma}_\ell s} \middle| \begin{array}{c} 2,1 \\ \frac{3}{2},2 \end{array}\right] = -\frac{\bar{\gamma}_\ell}{(1+2\bar{\gamma}_\ell s)^{3/2}},$$ where $\mathrm{G}_{p,q}^{m,n}[\cdot]$ and $\mathrm{H}_{p,q}^{m,n}[\cdot]$ represent the Meijer's G function [7, Eq. (8.2.1/1)] and Fox's H function [7, Eq. (8.3.1/1)], respectively. Note that one-sided Gaussian fading coincides with the worst-case fading or equivalently, the largest amount of fading (AoF) for all Gaussian-based fading distributions. |
| **Exponential** [1, Eq. (2.7)] $$p_{\gamma_\ell}(\gamma) = \frac{1}{\bar{\gamma}_\ell}\exp\left(-\frac{\gamma}{\bar{\gamma}_\ell}\right) = \frac{1}{\bar{\gamma}_\ell}\mathrm{H}_{0,1}^{1,0}\left[\frac{\gamma}{\bar{\gamma}_\ell} \middle| \begin{array}{c} \text{---} \\ (0,1) \end{array}\right],$$ where $\bar{\gamma}_\ell$ is the average power (i.e., $\bar{\gamma}_\ell \geq 0$). | $$\mathcal{M}_{\gamma_\ell}(s) = \mathrm{H}_{1,1}^{1,1}\left[\frac{1}{s\,\bar{\gamma}_\ell} \middle| \begin{array}{c} (1,1) \\ (1,1) \end{array}\right] = \mathrm{G}_{1,1}^{1,1}\left[\frac{1}{s\,\bar{\gamma}_\ell} \middle| \begin{array}{c} 1 \\ 1 \end{array}\right] = \frac{1}{1+\bar{\gamma}_\ell s},$$ $$\frac{\partial}{\partial s}\mathcal{M}_{\gamma_\ell}(s) = \bar{\gamma}_\ell\,\mathrm{H}_{2,2}^{2,1}\left[\frac{1}{s\,\bar{\gamma}_\ell} \middle| \begin{array}{c} (2,2),(1,1) \\ (2,1),(2,1) \end{array}\right] = \bar{\gamma}_\ell\,\mathrm{G}_{2,2}^{2,1}\left[\frac{1}{s\,\bar{\gamma}_\ell} \middle| \begin{array}{c} 2,1 \\ 2,2 \end{array}\right] = -\frac{\bar{\gamma}_\ell}{(1+s\,\bar{\gamma}_\ell)^2},$$ |
| **Gamma** [1, Eq. (2.21)] $$p_{\gamma_\ell}(\gamma) = \frac{1}{\Gamma(m_\ell)}\left(\frac{m_\ell}{\bar{\gamma}_\ell}\right)^{m_\ell}\gamma^{m_\ell-1}\exp\left(-\frac{m_\ell\gamma}{\bar{\gamma}_\ell}\right),$$ $$= \frac{m_\ell}{\Gamma(m_\ell)\bar{\gamma}_\ell}\mathrm{H}_{0,1}^{1,0}\left[\frac{m_\ell}{\bar{\gamma}_\ell}\gamma \middle| \begin{array}{c} \text{---} \\ (m_\ell-1,1) \end{array}\right],$$ where $\bar{\gamma}_\ell$ is the average power, and where $m_\ell$ $(0.5 \leq m_\ell)$ denotes the fading figure. Moreover, $\Gamma(\cdot)$ is the Gamma function [8, Sec. 8.31]. | $$\mathcal{M}_{\gamma_\ell}(s) = \frac{1}{\Gamma(m_\ell)}\mathrm{H}_{1,1}^{1,1}\left[\frac{m_\ell}{\bar{\gamma}_\ell s} \middle| \begin{array}{c} (1,1) \\ (m_\ell,1) \end{array}\right] = \frac{\mathrm{G}_{1,1}^{1,1}\left[\frac{m_\ell}{\bar{\gamma}_\ell s} \middle| \begin{array}{c} 1 \\ m_\ell \end{array}\right]}{\Gamma(m_\ell)} = \left(1+\frac{\bar{\gamma}_\ell}{m_\ell}s\right)^{-m_\ell},$$ $$\frac{\partial}{\partial s}\mathcal{M}_{\gamma_\ell}(s) = \frac{1}{\Gamma(m_\ell+1)}\mathrm{H}_{2,2}^{2,1}\left[\frac{m_\ell}{\bar{\gamma}_\ell s} \middle| \begin{array}{c} (2,1),(1,1) \\ (m_\ell+1,1),(2,1) \end{array}\right] = \frac{\mathrm{G}_{2,2}^{2,1}\left[\frac{m_\ell}{\bar{\gamma}_\ell s} \middle| \begin{array}{c} 2,1 \\ m_\ell+1,2 \end{array}\right]}{\Gamma(m_\ell+1)} = -\bar{\gamma}_\ell\left(1+\frac{\bar{\gamma}_\ell}{m_\ell}s\right)^{-m_\ell-1},$$ Note that the Nakagami-$m$ distribution spans via the $m$ parameter the widest range of amount of fading (AoF) among all the multipath distributions [1]. As such, Nakagami-$q$ (Hoyt) and Nakagami-$n$ (Rice) can also be closely approximated by Nakagami-$m$ distribution [1, Eq. (2.25)], [1, Eq. (2.26)]. |
| **Weibull** [1, Eq. (2.27)] $$p_{\gamma_\ell}(r) = \xi_\ell\left(\frac{\omega_\ell}{\bar{\gamma}_\ell}\right)^{\xi_\ell} r^{\xi_\ell-1}\exp\left(-\left(\frac{\omega_\ell}{\bar{\gamma}_\ell}\right)^{\xi_\ell}r^{\xi_\ell}\right),$$ $$= \frac{\omega_\ell}{\bar{\gamma}_\ell}\mathrm{H}_{0,1}^{1,0}\left[\frac{\omega_\ell}{\bar{\gamma}_\ell}\gamma \middle| \begin{array}{c} \text{---} \\ (1-1/\xi_\ell,1/\xi_\ell) \end{array}\right],$$ where $\omega_\ell = \Gamma(1+1/\xi_\ell)$ and where $\xi_\ell$ $(0 < \xi_\ell)$ denotes the fading shaping factor. Moreover, $\bar{\gamma}_\ell$ is the average power. | $$\mathcal{M}_{\gamma_\ell}(s) = \mathrm{H}_{1,1}^{1,1}\left[\frac{\omega_\ell}{\Omega_{s\ell} s} \middle| \begin{array}{c} (1,1) \\ (1,\frac{1}{\xi_\ell}) \end{array}\right] = \sqrt{\frac{4kl}{(2\pi)^{2k+2l-2}}}\mathrm{G}_{2k,2l}^{2l,2k}\left[\frac{\omega_\ell^{2k}(2k)^{2k}}{s^{2k}\Omega_{s\ell}^{2k}(2l)^{2l}} \middle| \begin{array}{c} -\Xi_{(2k)}^{(-2k)} \\ \Xi_{(2l)}^{(1)} \end{array}\right],$$ $$\frac{\partial}{\partial s}\mathcal{M}_{\gamma_\ell}(s) = \frac{\Omega_{s\ell}}{\omega_\ell}\mathrm{H}_{2,2}^{2,1}\left[\frac{\omega_\ell}{\Omega_{s\ell} s} \middle| \begin{array}{c} (2,1),(1,1) \\ (1+\frac{1}{\xi_\ell},\frac{1}{\xi_\ell}),(2,1) \end{array}\right] = \frac{\sqrt{16k^3 l}}{\sqrt{(2\pi)^{2k+2l-2}}\,s}\mathrm{G}_{2k+1,2l+1}^{2l+1,2k}\left[\frac{\omega_\ell^{2k}(2k)^{2k}}{s^{2k}\Omega_{s\ell}^{2k}(2l)^{2l}pl} \middle| \begin{array}{c} -\Xi_{(2k)}^{(-2k)},0 \\ \Xi_{(2l)}^{(1)},1 \end{array}\right],$$ where the Meijer's G representations are given for the rational value of the fading shaping factor $\xi_{s\ell}$ (that is, we let $\xi_{s\ell} = k/l$, where $k$, and $l$ are arbitrary positive integers.) through the medium of algebraic manipulations utilizing [7, Eq. (8.3.2.22)]. In addition, the coefficient $\Xi_{(n)}^{(x)}$ of the Meijer's G function is a set of coefficients such that it is defined as $\Xi_{(n)}^{(x)} \equiv \frac{x}{n}, \frac{x+1}{n}, \ldots, \frac{x+n-1}{n}$ with $x \in \mathbb{C}$ and $n \in \mathbb{N}$. |

TABLE III
MGFs OF SOME WELL-KNOWN FADING CHANNEL MODELS

| **Instantaneous SNR Distribution, i.e., $p_{\gamma_\ell}(\gamma)$** | **MGF $\mathcal{M}_{\gamma_\ell}(s)$ and its derivative $\frac{\partial}{\partial s}\mathcal{M}_{\gamma_\ell}(s)$** |
|---|---|
| **Hyper-Gamma** [22, Eq. (3)] $$p_{\gamma_\ell}(\gamma) = \sum_{k=1}^{K} \frac{\xi_{\ell k}}{\Gamma(m_{\ell k})} \left(\frac{m_{\ell k}}{\bar{\gamma}_{\ell k}}\right)^{m_{\ell k}} \gamma^{m_{\ell k}-1} e^{-\frac{m_{\ell k}}{\bar{\gamma}_{\ell k}}\gamma},$$ $$= \sum_{k=1}^{K} \frac{\xi_{\ell k} m_{\ell k}}{\Gamma(m_{\ell k})\bar{\gamma}_{\ell k}} \mathrm{H}_{0,1}^{1,0}\!\left[\frac{m_{\ell k}}{\bar{\gamma}_{\ell k}}\gamma \,\Big|\, \begin{matrix}-\\(m_{\ell k}-1,1)\end{matrix}\right],$$ where $m_{\ell k}$ ($0.5 \le m_{\ell k}$) is the fading figure, $\bar{\gamma}_{\ell k}$ ($0<\bar{\gamma}_{\ell k}$) is the average power, and $\xi_{\ell k}$ ($0<\xi_{\ell k}$) is the accruing factor, of the $k$th fading environment. | $$\mathcal{M}_{\gamma_\ell}(s) = \sum_{k=1}^{K} \frac{\xi_{\ell k} \mathrm{H}_{1,1}^{1,1}\!\left[\frac{m_{\ell k}}{s\bar{\gamma}_{\ell k}}\,\Big|\,\begin{matrix}(1,1)\\(m_{\ell k},1)\end{matrix}\right]}{\Gamma(m_{\ell k})} = \sum_{k=1}^{K} \frac{\xi_{\ell k} \mathrm{G}_{1,1}^{1,1}\!\left[\frac{m_{\ell k}}{s\bar{\gamma}_{\ell k}}\,\Big|\,\begin{matrix}1\\m_{\ell k}\end{matrix}\right]}{\Gamma(m_{\ell k})} = \sum_{k=1}^{K} \xi_{\ell k}\left(1 + \frac{\bar{\gamma}_{\ell k}}{m_{\ell k}}s\right)^{-m_{\ell k}},$$ $$\frac{\partial}{\partial s}\mathcal{M}_{\gamma_\ell}(s) = \sum_{k=1}^{K} \frac{\xi_{\ell k}\mathrm{H}_{2,2}^{2,1}\!\left[\frac{m_{\ell k}}{s\bar{\gamma}_{\ell k}}\,\Big|\,\begin{matrix}(1,1),(0,1)\\(m_{\ell k},1),(1,1)\end{matrix}\right]}{s\,\Gamma(m_{\ell k})} = \sum_{k=1}^{K}\frac{\xi_{\ell k}\mathrm{G}_{2,2}^{2,1}\!\left[\frac{m_{\ell k}}{s\bar{\gamma}_{\ell k}}\,\Big|\,\begin{matrix}1,0\\m_{\ell k},1\end{matrix}\right]}{s\,\Gamma(m_{\ell k})} = -\sum_{k=1}^{K}\xi_{\ell k}\bar{\gamma}_{\ell k}\left(1+\frac{\bar{\gamma}_{\ell k}}{m_{\ell k}}s\right)^{-m_{\ell k}-1},$$ where $\Gamma(\cdot)$ is the Gamma function [6, Eq. (6.1.1)]. In addition, It may be useful to notice that the sum of the accruing probabilities $\xi_{\ell k}$, $k \in \{1,2,\ldots,K\}$ of $K$ possible fading environments is unit such that $\sum_{k=1}^{K} \xi_{\ell k} = 1$. |
| **Power of Nakagami-$q$ (Hoyt)** [1, Eq. (2.11)] $$p_{\gamma_\ell}(\gamma) = \frac{1+q_\ell^2}{2q_\ell \bar{\gamma}_\ell} e^{-\frac{(1+q_\ell^2)^2}{4q_\ell^2\bar{\gamma}_\ell}\gamma} I_0\!\left(\frac{1-q_\ell^4}{4q_\ell^2\bar{\gamma}_\ell}\gamma\right),$$ $$= \lim_{K\to\infty}\sum_{k=0}^{K}\frac{\Phi_k m_k \mathrm{H}_{0,1}^{1,0}\!\left[\frac{m_k}{\Omega_k}\gamma\,\Big|\,\begin{matrix}-\\(m_k-1,1)\end{matrix}\right]}{\Gamma(m_k)\Omega_k},$$ where $q_\ell$ ($0<q_\ell<1$) is the Nakagami-$q$ fading parameter and $\bar{\gamma}_\ell$ ($0<\bar{\gamma}_\ell$) is the average power. In addition, $I_0(\cdot)$ is the zeroth order modified Bessel function of the first kind [6, Eq. (9.6.20)]. | $$\mathcal{M}_{\gamma_\ell}(s) = \lim_{K\to\infty}\sum_{k=0}^{K}\frac{\Phi_k}{\Gamma(m_k)}\mathrm{H}_{1,1}^{1,1}\!\left[\frac{m_k}{s\,\Omega_k}\,\Big|\,\begin{matrix}(1,1)\\(m_k,1)\end{matrix}\right] = \left(1+2\bar{\gamma}_\ell s+\frac{(2\bar{\gamma}_\ell s)^2 q_\ell^2}{(1+q_\ell^2)^2}\right)^{-\frac{1}{2}},$$ $$\frac{\partial}{\partial s}\mathcal{M}_{\gamma_\ell}(s) = \lim_{K\to\infty}\sum_{k=0}^{K}\frac{\Phi_k}{s\,\Gamma(m_k)}\mathrm{H}_{2,2}^{2,1}\!\left[\frac{m_k}{s\,\Omega_k}\,\Big|\,\begin{matrix}(1,1),(0,1)\\(m_k,1),(1,1)\end{matrix}\right] = -\frac{\bar{\gamma}_\ell\left(1+\frac{4q_\ell^2\bar{\gamma}_\ell s}{(1+q_\ell^2)^2}\right)}{\left(1+2\bar{\gamma}_\ell s+\frac{(2\bar{\gamma}_\ell s)^2 q_\ell^2}{(1+q_\ell^2)^2}\right)^{\frac{3}{2}}},$$ where $m_k,\Omega_k$ are defined as $m_k = 2k+1$ and $\Omega_k = 4(2k+1)q_\ell^2\bar{\gamma}_\ell/(1+q_\ell^2)^2$, respectively. In addition, the weighting coefficients $\{\Phi_k\}$ are given by $\Phi_k = \frac{2q_\ell}{\sqrt{\pi}(1+q_\ell^2)}\frac{\Gamma(k+\frac{1}{2})}{\Gamma(k+1)}\left(\frac{1-q_\ell^2}{1+q_\ell^2}\right)^{2k}$ for all $k\in\mathbb{N}$. It may be useful to notice that the series expression of the MGF for the Nakagami-$q$ (Hoyt) and its derivative are converging very fast such that 10 summation terms is generally enough. |
| **Power of Nakagami-$n$ (Rice)** [1, Eq. (2.16)] $$p_{\gamma_\ell}(\gamma) = \frac{(1+n_\ell^2)e^{-n_\ell^2}}{\bar{\gamma}_\ell}e^{-\frac{(1+n_\ell^2)}{\bar{\gamma}_\ell}\gamma} I_0\!\left(2n_\ell\sqrt{\frac{1+n_\ell^2}{\bar{\gamma}_\ell}\gamma}\right),$$ $$= \lim_{K\to\infty}\sum_{k=0}^{K}\frac{\Psi_k m_k \mathrm{H}_{0,1}^{1,0}\!\left[\frac{m_k}{\Omega_k}\gamma\,\Big|\,\begin{matrix}-\\(m_k-1,1)\end{matrix}\right]}{\Gamma(m_k)\Omega_k},$$ where $n_\ell$ ($0<n_\ell$) and $\bar{\gamma}_\ell$ ($0<\bar{\gamma}_\ell$) are the LOS figure and average power, respectively. | $$\mathcal{M}_{\gamma_\ell}(s) = \lim_{K\to\infty}\sum_{k=0}^{K}\frac{\Psi_k}{\Gamma(m_k)}\mathrm{H}_{1,1}^{1,1}\!\left[\frac{m_k}{s\,\Omega_k}\,\Big|\,\begin{matrix}(1,1)\\(m_k,1)\end{matrix}\right] = \frac{1+n_\ell^2}{(1+n_\ell^2)+\bar{\gamma}_\ell s}\exp\!\left(-\frac{n_\ell^2\bar{\gamma}_\ell s}{(1+n_\ell^2)+\bar{\gamma}_\ell s}\right),$$ $$\frac{\partial}{\partial s}\mathcal{M}_{\gamma_\ell}(s) = \lim_{K\to\infty}\sum_{k=0}^{K}\frac{\Psi_k}{s\,\Gamma(m_k)}\mathrm{H}_{2,2}^{2,1}\!\left[\frac{m_k}{s\,\Omega_k}\,\Big|\,\begin{matrix}(1,1),(0,1)\\(m_k,1),(1,1)\end{matrix}\right] = -\bar{\gamma}_\ell\frac{1+\frac{\bar{\gamma}_\ell s}{(1+n_\ell^2)^2}}{\left(1+\frac{\bar{\gamma}_\ell s}{1+n_\ell^2}\right)^3}\exp\!\left(-\frac{n_\ell^2\bar{\gamma}_\ell s}{(1+n_\ell^2)+\bar{\gamma}_\ell s}\right),$$ where $m_k$ and $\Omega_k$ are defined as $m_k = k+1$ and $\Omega_k = (k+1)\frac{\bar{\gamma}_\ell}{1+n_\ell^2}$, respectively. In addition, the weighting coefficients $\Psi_k$ are given by $\Psi_k = n_\ell^{2k}\exp(-n_\ell^2)/\Gamma(k+1)$. It may be useful to notice that the line-of-sight (LOS) figure i.e. $n_\ell$ is related to the Rician $K_\ell$ factor by $K_\ell = n_\ell^2$ which corresponds to the ratio of the power of the LOS (specular) component to the average power of the scattered component. |

TABLE IV
MGFs OF SOME WELL-KNOWN FADING CHANNEL MODELS

| Instantaneous SNR Distribution, i.e., $p_{\gamma_\ell}(\gamma)$ | MGF $\mathcal{M}_{\gamma_\ell}(s)$ and its derivative $\frac{\partial}{\partial s}\mathcal{M}_{\gamma_\ell}(s)$ |
|---|---|
| **Maxwell** [1, Eq. (2.53)] $$p_{\gamma_\ell}(\gamma) = \sqrt{\frac{27\,\gamma}{2\pi\,\bar{\gamma}_\ell^3}}\,e^{-\frac{3}{2\bar{\gamma}_\ell}\gamma} = \frac{3\,\mathrm{H}_{0,1}^{1,0}\!\left[\frac{3\gamma}{2\bar{\gamma}_\ell}\,\middle|\,\overline{(\frac{1}{2},1)}\right]}{\sqrt{\pi}\,\bar{\gamma}_\ell},$$ where $\bar{\gamma}_\ell$ is the average power. | $$\mathcal{M}_{\gamma_\ell}(s) = \frac{2}{\sqrt{\pi}}\mathrm{H}_{1,1}^{1,1}\!\left[\frac{3}{2\bar{\gamma}_\ell\,s}\,\middle|\,{(1,1) \atop (\frac{3}{2},1)}\right] = \frac{2}{\sqrt{\pi}}\mathrm{G}_{1,1}^{1,1}\!\left[\frac{3}{2\bar{\gamma}_\ell\,s}\,\middle|\,{1 \atop \frac{3}{2}}\right] = \frac{3\sqrt{3}}{(3+2\bar{\gamma}_\ell\,s)^{3/2}},$$ $$\frac{\partial}{\partial s}\mathcal{M}_{\gamma_\ell}(s) = \frac{2}{\sqrt{\pi}\,s}\mathrm{H}_{2,2}^{2,1}\!\left[\frac{3}{2\bar{\gamma}_\ell\,s}\,\middle|\,{(1,1),(0,1) \atop (\frac{3}{2},1),(1,1)}\right] = -\frac{9\sqrt{3}\,\bar{\gamma}_\ell}{(3+2\bar{\gamma}_\ell\,s)^{5/2}},$$ |
| **Lognormal** [1, Eq. (2.53)] $$p_{\gamma_\ell}(\gamma) = \frac{\xi}{\sqrt{2\pi}\sigma_\ell\gamma}\,e^{-\frac{(10\log_{10}(\gamma)-\mu_\ell)^2}{2\sigma_\ell^2}},$$ $$= \frac{1}{\sqrt{\pi}}\sum_{k=1}^{K}\frac{w_k}{\omega_k}\mathrm{H}_{0,0}^{0,0}\!\left[\frac{\gamma}{\omega_k}\,\middle|\,\overline{\phantom{xx}}\right],$$ where $\mu_\ell$(dB) and $\sigma_\ell$(dB) are the mean and the standard deviation of $\gamma_\ell$. | $$\mathcal{M}_{\gamma_\ell}(s) = \frac{1}{\sqrt{\pi}}\sum_{k=1}^{K}w_k\mathrm{H}_{0,1}^{1,0}\!\left[\omega_k\,s\,\middle|\,\overline{(0,1)}\right] = \frac{1}{\sqrt{\pi}}\sum_{k=1}^{K}w_k\exp(-\omega_k\,s),$$ $$\frac{\partial}{\partial s}\mathcal{M}_{\gamma_\ell}(s) = -\frac{1}{\sqrt{\pi}}\sum_{k=1}^{K}w_k\omega_k\mathrm{H}_{0,1}^{1,0}\!\left[\omega_k\,s\,\middle|\,\overline{(0,1)}\right] = -\frac{1}{\sqrt{\pi}}\sum_{k=1}^{K}w_k\omega_k\exp(-\omega_k\,s),$$ where $\omega_k$ is defined as $\omega_k = 10^{(\sqrt{2}\sigma_\ell u_k+\mu_\ell)/10}$ such that for $k\in\{1,2,\ldots,K\}$, $\{w_k\}$ and $\{u_k\}$ are the weight factors and the zeros (abscissas) of the $K$-order Hermite polynomial [6, Table 25.10]. |
| **Power of K-Distribution** [1, Eq. (2.15)] $$p_{\gamma_\ell}(\gamma) = \frac{2\left(\frac{m_{s\ell}}{\bar{\gamma}_{s\ell}}\right)^{\frac{m_{s\ell}+1}{2}}\gamma^{\frac{m_{s\ell}-1}{2}}}{\Gamma(m_{s\ell})}K_{m_{s\ell}-1}\!\left(2\sqrt{\frac{m_{s\ell}\gamma}{\bar{\gamma}_{s\ell}}}\right),$$ $$= \frac{m_{s\ell}\,\mathrm{H}_{0,2}^{2,0}\!\left[\frac{m_{s\ell}}{\bar{\gamma}_{s\ell}}s\,\middle|\,\overline{(0,1),(m_{s\ell}-1,1)}\right]}{\Gamma(m_{s\ell})\bar{\gamma}_{s\ell}},$$ | $$\mathcal{M}_{\gamma_\ell}(s) = \frac{1}{\Gamma(m_{s\ell})}\mathrm{H}_{1,2}^{2,1}\!\left[\frac{m_{s\ell}}{\bar{\gamma}_{s\ell}\,s}\,\middle|\,{(1,1) \atop (1,1),(m_{s\ell},1)}\right] = \frac{1}{\Gamma(m_{s\ell})}\mathrm{G}_{1,2}^{2,1}\!\left[\frac{m_{s\ell}}{\bar{\gamma}_{s\ell}\,s}\,\middle|\,{1 \atop 1,m_{s\ell}}\right],$$ $$\frac{\partial}{\partial s}\mathcal{M}_{\gamma_\ell}(s) = \frac{1}{\Gamma(m_{s\ell})\,s}\mathrm{H}_{2,3}^{3,1}\!\left[\frac{m_{s\ell}}{\bar{\gamma}_{s\ell}\,s}\,\middle|\,{(1,1),(0,1) \atop (1,1),(1,1),(m_{s\ell},1)}\right] = \frac{1}{\Gamma(m_{s\ell})\,s}\mathrm{G}_{2,3}^{3,1}\!\left[\frac{m_{s\ell}}{\bar{\gamma}_{s\ell}\,s}\,\middle|\,{1,0 \atop 1,1,m_{s\ell}}\right],$$ where $m_{s\ell}$ ($\frac{1}{2}\le m_{s\ell}$) denotes the shadowing severity, and $\bar{\gamma}_{s\ell}$ ($0<\bar{\gamma}_{s\ell}$) represents the average power. In addition, $K_n(\cdot)$ is the $n$th order modified Bessel function of the second kind [6, Eq. (9.6.24)]. |
| **Power of Generalized-K** [23, Eq. (5)] $$p_{\gamma_\ell}(\gamma) = \frac{2\left(\frac{m_{s\ell}m_\ell}{\bar{\gamma}_{s\ell}}\right)^{\frac{\phi_\ell}{2}}\gamma^{\frac{\phi_\ell}{2}-1}}{\Gamma(m_{s\ell})}K_{\psi_\ell}\!\left(2\sqrt{\frac{m_{s\ell}m_\ell\gamma}{\bar{\gamma}_{s\ell}}}\right),$$ $$= \frac{m_\ell m_{s\ell}\,\mathrm{H}_{0,2}^{2,0}\!\left[\frac{m_\ell m_{s\ell}}{\bar{\gamma}_{s\ell}}\gamma\,\middle|\,\overline{(m_\ell-1,1),(m_{s\ell}-1,1)}\right]}{\Gamma(m_\ell)\Gamma(m_{s\ell})\bar{\gamma}_{s\ell}},$$ | $$\mathcal{M}_{\gamma_\ell}(s) = \frac{1}{\Gamma(m_\ell)\Gamma(m_{s\ell})}\mathrm{H}_{1,2}^{2,1}\!\left[\frac{m_{s\ell}m_\ell}{\bar{\gamma}_{s\ell}\,s}\,\middle|\,{(1,1) \atop (m_\ell,1),(m_{s\ell},1)}\right] = \frac{1}{\Gamma(m_\ell)\Gamma(m_{s\ell})}\mathrm{G}_{1,2}^{2,1}\!\left[\frac{m_{s\ell}m_\ell}{\bar{\gamma}_{s\ell}\,s}\,\middle|\,{1 \atop m_\ell,m_{s\ell}}\right],$$ $$\frac{\partial}{\partial s}\mathcal{M}_{\gamma_\ell}(s) = \frac{1}{\Gamma(m_\ell)\Gamma(m_{s\ell})\,s}\mathrm{H}_{2,3}^{3,1}\!\left[\frac{m_{s\ell}m_\ell}{\bar{\gamma}_{s\ell}\,s}\,\middle|\,{(1,1),(0,1) \atop (m_\ell,1),(m_{s\ell},1),(1,1)}\right] = \frac{\mathrm{G}_{2,3}^{3,1}\!\left[\frac{m_{s\ell}m_\ell}{\bar{\gamma}_{s\ell}\,s}\,\middle|\,{1,0 \atop m_\ell,m_{s\ell},1}\right]}{\Gamma(m_\ell)\Gamma(m_{s\ell})\,s},$$ where $\phi_\ell = m_{s\ell}+m_\ell$ and $\psi_\ell = m_{s\ell}-m_\ell$, and where the parameters $m_\ell$ ($0.5\le m_\ell$) and $m_{s\ell}$ ($0.5\le m_{s\ell}$) represent the fading figure (diversity severity / order) and the shadowing severity, respectively. $\bar{\gamma}_{s\ell}$ ($0<\bar{\gamma}_{s\ell}$) represents the average power. |

TABLE V
MGFs OF SOME WELL-KNOWN FADING CHANNEL MODELS

| **Instantaneous SNR Distribution, i.e., $p_{\gamma_\ell}(\gamma)$** | **MGF $\mathcal{M}_{\gamma_\ell}(s)$ and its derivative $\frac{\partial}{\partial s}\mathcal{M}_{\gamma_\ell}(s)$** |
|---|---|
| **Generalized Gamma** [24] $$p_{\gamma_\ell}(\gamma) = \frac{\xi_\ell \left(\frac{\beta_\ell}{\bar{\gamma}_\ell}\right)^{m_\ell \xi_\ell} \gamma^{m_\ell \xi_\ell - 1}}{\Gamma(m_\ell)} \exp\left(-\left(\frac{\beta_\ell}{\bar{\gamma}_\ell}\gamma\right)^{\xi_\ell}\right)$$ $$= \frac{\beta_\ell}{\Gamma(m_\ell)\bar{\gamma}_\ell} H_{0,1}^{1,0}\left[\frac{\beta_\ell}{\bar{\gamma}_\ell}\gamma \,\middle|\, \genfrac{}{}{0pt}{}{-}{(m_\ell - \frac{1}{\xi_\ell}, \frac{1}{\xi_\ell})}\right]$$ | $$\mathcal{M}_{\gamma_\ell}(s) = \frac{H_{1,1}^{1,1}\left[\frac{\beta_\ell}{\bar{\gamma}_\ell s} \,\middle|\, \genfrac{}{}{0pt}{}{(1,1)}{(m_\ell, \frac{1}{\xi_\ell})}\right]}{\Gamma(m_\ell)} = \frac{2\pi\, l^{m_\ell} k}{\sqrt{(2\pi)^{k+l}\, kl}\,\Gamma(m_\ell)} G_{k,l}^{l,k}\left[\frac{k^k}{l^l}\left(\frac{\beta_\ell}{\bar{\gamma}_\ell s}\right)^k \,\middle|\, \genfrac{}{}{0pt}{}{\Xi_{(1)}^{(k)}}{\Xi_{(m_\ell)}^{(l)}}\right],$$ $$\frac{\partial}{\partial s}\mathcal{M}_{\gamma_\ell}(s) = \frac{H_{2,2}^{2,1}\left[\frac{\beta_\ell}{\bar{\gamma}_\ell s} \,\middle|\, \genfrac{}{}{0pt}{}{(1,1),(0,1)}{(m_\ell, \frac{1}{\xi_\ell}),(1,1)}\right]}{\Gamma(m_\ell)\, s} = \frac{2\pi\, l^{m_\ell} k^2}{\sqrt{(2\pi)^{k+l}\, kl}\,\Gamma(m_\ell)\, s} G_{k+1,l+1}^{l+1,k}\left[\frac{k^k}{l^l}\left(\frac{\beta_\ell}{\bar{\gamma}_\ell s}\right)^k \,\middle|\, \genfrac{}{}{0pt}{}{\Xi_{(1)}^{(k)},0}{\Xi_{(m_\ell)}^{(l)},1}\right],$$ where the parameters $m_\ell$ ($0.5 \le m_\ell < \infty$) and $\xi_\ell$ ($0 \le \xi_\ell < \infty$) represent the fading figure (diversity severity / order) and the fading shaping factor, respectively, while $\bar{\gamma}_\ell$ ($0 \le \bar{\gamma}_\ell < \infty$) is the average power. In addition, the parameter $\beta_\ell$ is defined as $\beta_\ell = \Gamma(m_\ell + 1/\xi_\ell)/\Gamma(m_\ell)$, and referring the coefficients of the Meijer's G function, $\Xi_{(n)}^{(x)}$ is a set of coefficients such that it is defined as $\Xi_{(n)}^{(x)} \equiv \frac{x}{n}, \frac{x+1}{n}, \ldots, \frac{x+n-1}{n}$ with $x \in \mathbb{C}$ and $n \in \mathbb{N}$. |
| **Extended Generalized Gamma** [16] $$p_{\gamma_\ell}(\gamma) = \frac{\xi_\ell \left(\frac{\beta_\ell \beta_{s\ell}}{\bar{\gamma}_{s\ell}}\right)^{\xi_\ell m_\ell}}{\Gamma(m_\ell)\Gamma(m_{s\ell})}\gamma^{\xi_\ell m_\ell - 1} \times \Gamma\left(m_{s\ell} - m_\ell \frac{\xi_\ell}{\xi_{s\ell}}, 0, \left(\frac{\beta_\ell \beta_{s\ell}}{\bar{\gamma}_{s\ell}}\right)^{\xi_\ell}\gamma^{\xi_\ell}, \frac{\xi_\ell}{\xi_{s\ell}}\right)$$ $$= \frac{H_{0,2}^{2,0}\left[\frac{\beta_\ell \beta_{s\ell}}{\bar{\gamma}_{s\ell}}\gamma \,\middle|\, \genfrac{}{}{0pt}{}{-}{(m_\ell - \frac{1}{\xi_\ell}, \frac{1}{\xi_\ell}),(m_{s\ell} - \frac{1}{\xi_{s\ell}}, \frac{1}{\xi_{s\ell}})}\right]}{\frac{\Gamma(m_\ell)\Gamma(m_{s\ell})\bar{\gamma}_{s\ell}}{\beta_\ell \beta_{s\ell}}},$$ | $$\mathcal{M}_{\gamma_\ell}(s) = \frac{H_{1,2}^{2,1}\left[\frac{\beta_\ell \beta_{s\ell}}{\bar{\gamma}_{s\ell}\, s} \,\middle|\, \genfrac{}{}{0pt}{}{(1,1)}{(m_\ell, \frac{1}{\xi_\ell}),(m_{s\ell}, \frac{1}{\xi_{s\ell}})}\right]}{\Gamma(m_\ell)\Gamma(m_{s\ell})} = \frac{\Phi_\ell\, G_{kk_s,\, kl_s+lk_s}^{kl_s+lk_s,\, kk_s}\left[\left(\frac{\beta_\ell \beta_{s\ell}}{\Psi_\ell \bar{\gamma}_{s\ell}\, s}\right)^{kk_s} \,\middle|\, \genfrac{}{}{0pt}{}{\Xi_{(1)}^{(k_s k)}}{\Xi_{(m_\ell)}^{(l k_s)},\Xi_{(m_{s\ell})}^{(l_s k)}}\right]}{\Gamma(m_\ell)\Gamma(m_{s\ell})},$$ $$\frac{\partial}{\partial s}\mathcal{M}_{\gamma_\ell}(s) = \frac{H_{1,2}^{2,1}\left[\frac{\beta_\ell \beta_{s\ell}}{\bar{\gamma}_{s\ell}\, s} \,\middle|\, \genfrac{}{}{0pt}{}{(1,1)}{(m_\ell, \frac{1}{\xi_\ell}),(m_{s\ell}, \frac{1}{\xi_{s\ell}})}\right]}{\Gamma(m_\ell)\Gamma(m_{s\ell})} = \frac{\Phi_\ell\, kk_s\, G_{kk_s+1,\, kl_s+lk_s+1}^{kl_s+lk_s+1,\, kk_s}\left[\left(\frac{\beta_\ell \beta_{s\ell}}{\Psi_\ell \bar{\gamma}_{s\ell}\, s}\right)^{kk_s} \,\middle|\, \genfrac{}{}{0pt}{}{\Xi_{(1)}^{(k_s k)},0}{\Xi_{(m_\ell)}^{(l k_s)},\Xi_{(m_{s\ell})}^{(l_s k)},1}\right]}{\Gamma(m_\ell)\Gamma(m_{s\ell})\, s},$$ where $\Phi_\ell = 2\pi(lk_s)^{m_\ell}(l_s k)^{m_{s\ell}}/\sqrt{(2\pi)^{lk_s+l_s k+kk_s-1}ll_s}$ and $\Psi_\ell = (lk_s)^{\frac{1}{\xi_\ell}}(l_s k)^{\frac{1}{\xi_{s\ell}}}/(kk_s)$, and where the parameters $m_\ell$ ($0.5 \le m_\ell < \infty$) and $\xi_\ell$ ($0 \le \xi_\ell < \infty$) represent the fading figure (diversity severity / order) and the fading shaping factor, respectively, while $m_{s\ell}$ ($0.5 \le m_{s\ell} < \infty$) and $\xi_{s\ell}$ ($0 \le \xi_{s\ell} < \infty$) represent the shadowing severity and the shadowing shaping factor (inhomogeneity), respectively. In addition, the parameters $\beta_\ell$ and $\beta_{s\ell}$ are defined as $\beta_\ell = \Gamma(m_\ell + 1/\xi_\ell)/\Gamma(m_\ell)$ and $\beta_{s\ell} = \Gamma(m_{s\ell} + 1/\xi_{s\ell})/\Gamma(m_{s\ell})$, respectively, where $\Gamma(\cdot)$ is the Gamma function [6, Eq. (6.5.3)]. In addition, $\Gamma(\cdot,\cdot,\cdot,\cdot)$ is the extended incomplete Gamma function defined as $\Gamma(\alpha, x, b, \beta) = \int_x^\infty r^{\alpha-1}\exp(-r - br^{-\beta})\,dr$, where $\alpha, \beta, b \in \mathbb{C}$ and $x \in \mathbb{R}^+$ [17, Eq. (6.2)]. Referring the coefficients of the Meijer's G function, $\Xi_{(n)}^{(x)}$ is a set of coefficients such that it is defined as $\Xi_{(n)}^{(x)} \equiv \frac{x}{n}, \frac{x+1}{n}, \ldots, \frac{x+n-1}{n}$ with $x \in \mathbb{C}$ and $n \in \mathbb{N}$. |
| **Fox's H distribution** [18, Eq. (3.1)], [9] $$p_{\gamma_\ell}(\gamma) = \mathcal{K}_\ell H_{p,q}^{m,n}\left[\mathcal{G}_\ell \gamma \,\middle|\, \genfrac{}{}{0pt}{}{(a_i, \alpha_i)_{i=1,2,\ldots,p}}{(b_j, \beta_j)_{j=1,2,\ldots,q}}\right]$$ where $\mathcal{K}_\ell \in \mathbb{R}$ and $\mathcal{G}_\ell \in \mathbb{R}$ are such two numbers that $\int_0^\infty p_{\gamma_\ell}(\gamma)\,d\gamma = 1$. | $$\mathcal{M}_{\gamma_\ell}(s) = \frac{\mathcal{K}_\ell}{\mathcal{G}_\ell} H_{p+1,q}^{m,n+1}\left[\frac{\mathcal{G}_\ell}{s} \,\middle|\, \genfrac{}{}{0pt}{}{(1,1),(a_i + \alpha_i, \alpha_i)_{i=1,2,\ldots,p}}{(b_j + \beta_j, \beta_j)_{j=1,2,\ldots,q}}\right],$$ $$\frac{\partial}{\partial s}\mathcal{M}_{\gamma_\ell}(s) = -\frac{\mathcal{K}_\ell}{\mathcal{G}_\ell\, s} H_{p+1,q}^{m,n+1}\left[\frac{\mathcal{G}_\ell}{s} \,\middle|\, \genfrac{}{}{0pt}{}{(0,1),(a_i + \alpha_i, \alpha_i)_{i=1,2,\ldots,p}}{(b_j + \beta_j, \beta_j)_{j=1,2,\ldots,q}}\right],$$ where $\max_{i \in \{1,2,\ldots,m\}}\{-b_i/\beta_i\} < \min_{i \in \{1,2,\ldots,n\}}\{(1 - a_i)/\alpha_i\}$ where $a_i, b_i \in \mathbb{R}$ and $\alpha_i, \beta_i \in \mathbb{R}^+$. |

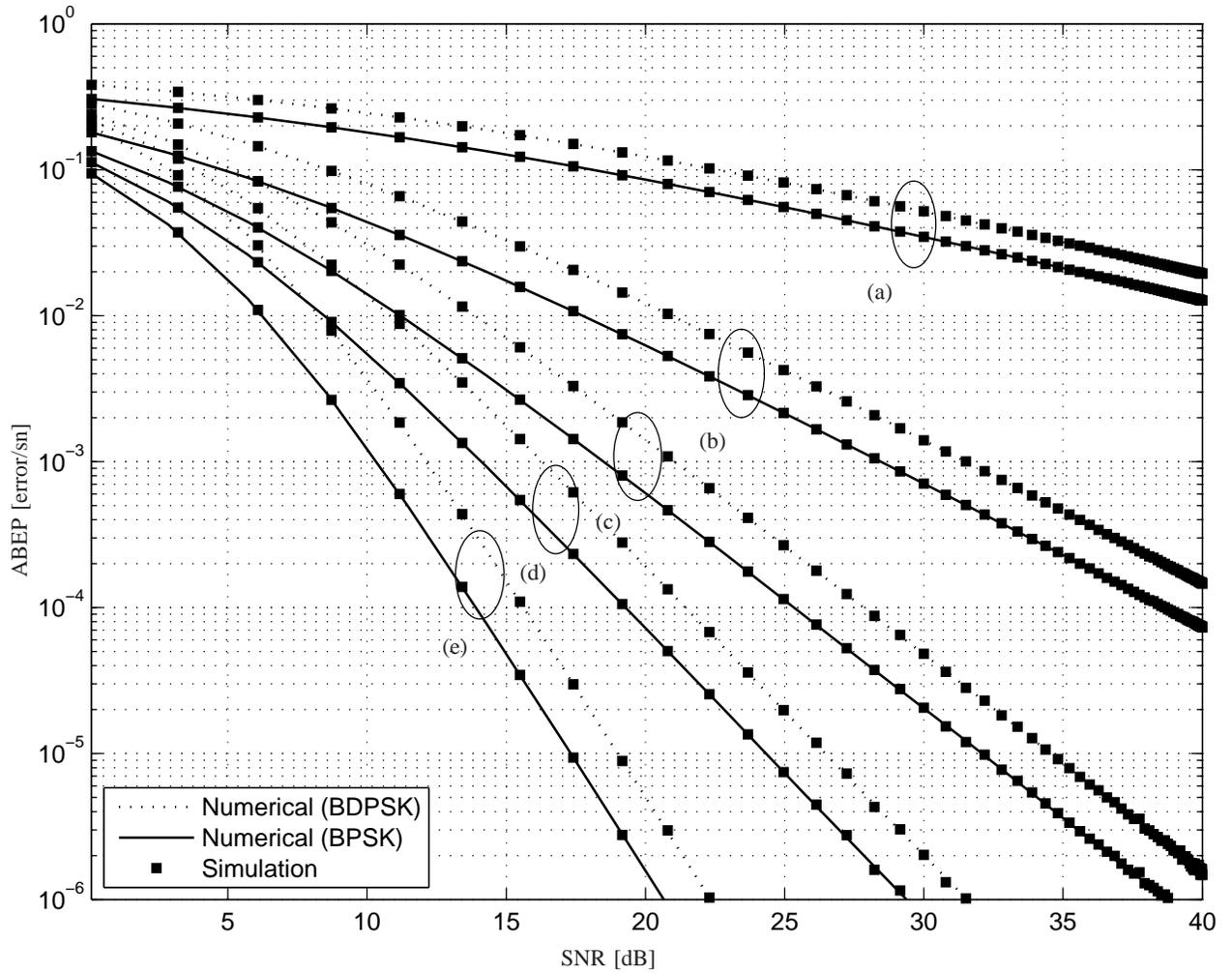

Fig. 1. ABEP performance comparison of BPSK and BDPSK binary modulations over generalized Gamma fading channel with the parameters: (a) $m_\ell = 2$, $\xi_\ell = 0.25$, (b) $m_\ell = 2$, $\xi_\ell = 0.5$, (c) $m_\ell = 2$, $\xi_\ell = 0.75$, (d) $m_\ell = 2$, $\xi_\ell = 1.0$ and (e) $m_\ell = 2$, $\xi_\ell = 1.5$. The number of samples is chosen as $10^7$ in the computer-based simulations.

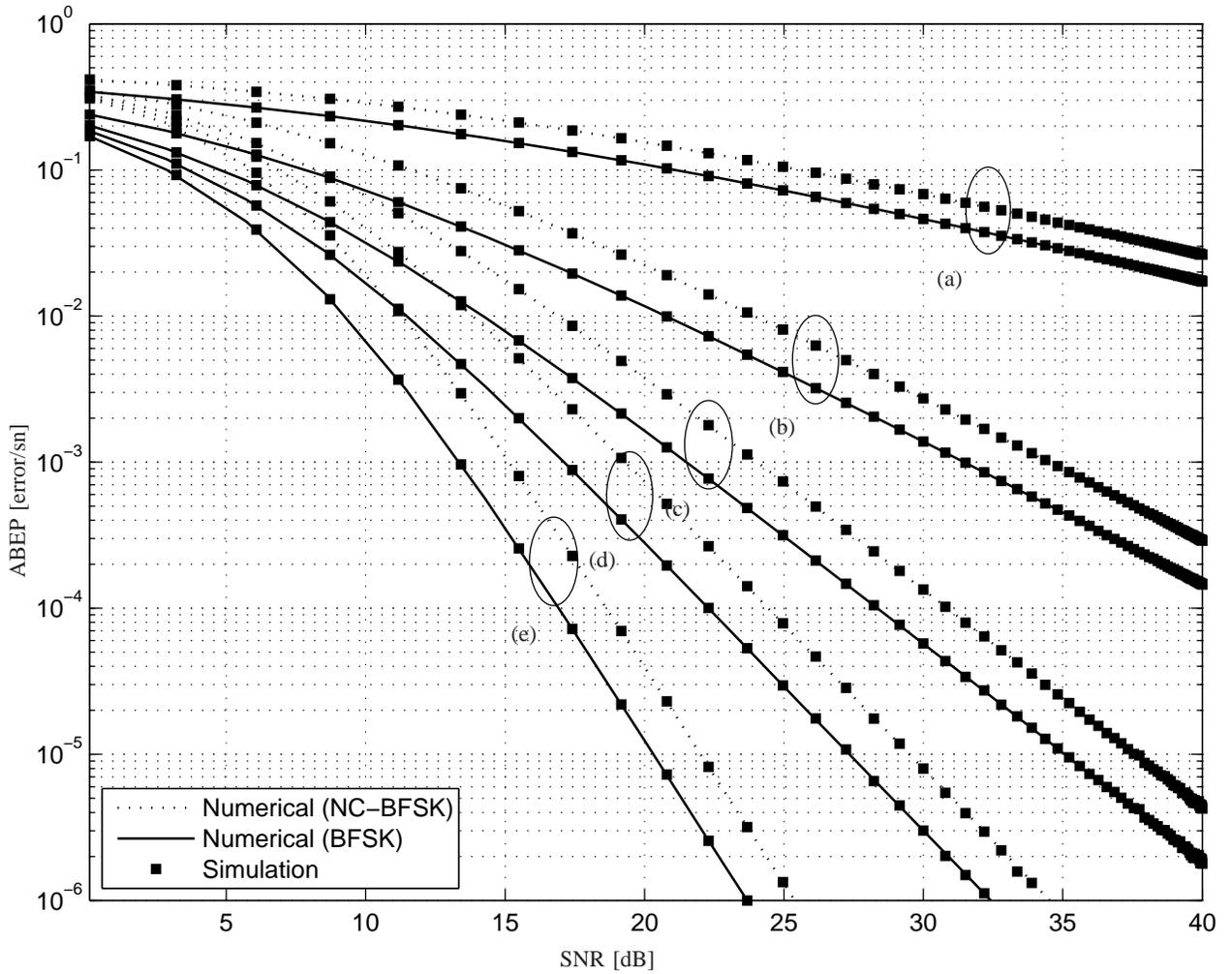

Fig. 2. ABEP performance comparison of BFSK and NC-FSK binary modulations over generalized Gamma fading channels with the parameters: (a) $m_\ell = 2$, $\xi_\ell = 0.25$, (b) $m_\ell = 2$, $\xi_\ell = 0.5$, (c) $m_\ell = 2$, $\xi_\ell = 0.75$, (d) $m_\ell = 2$, $\xi_\ell = 1.0$ and (e) $m_\ell = 2$, $\xi_\ell = 1.5$. The number of samples is chosen as $10^7$ in the computer-based simulations.

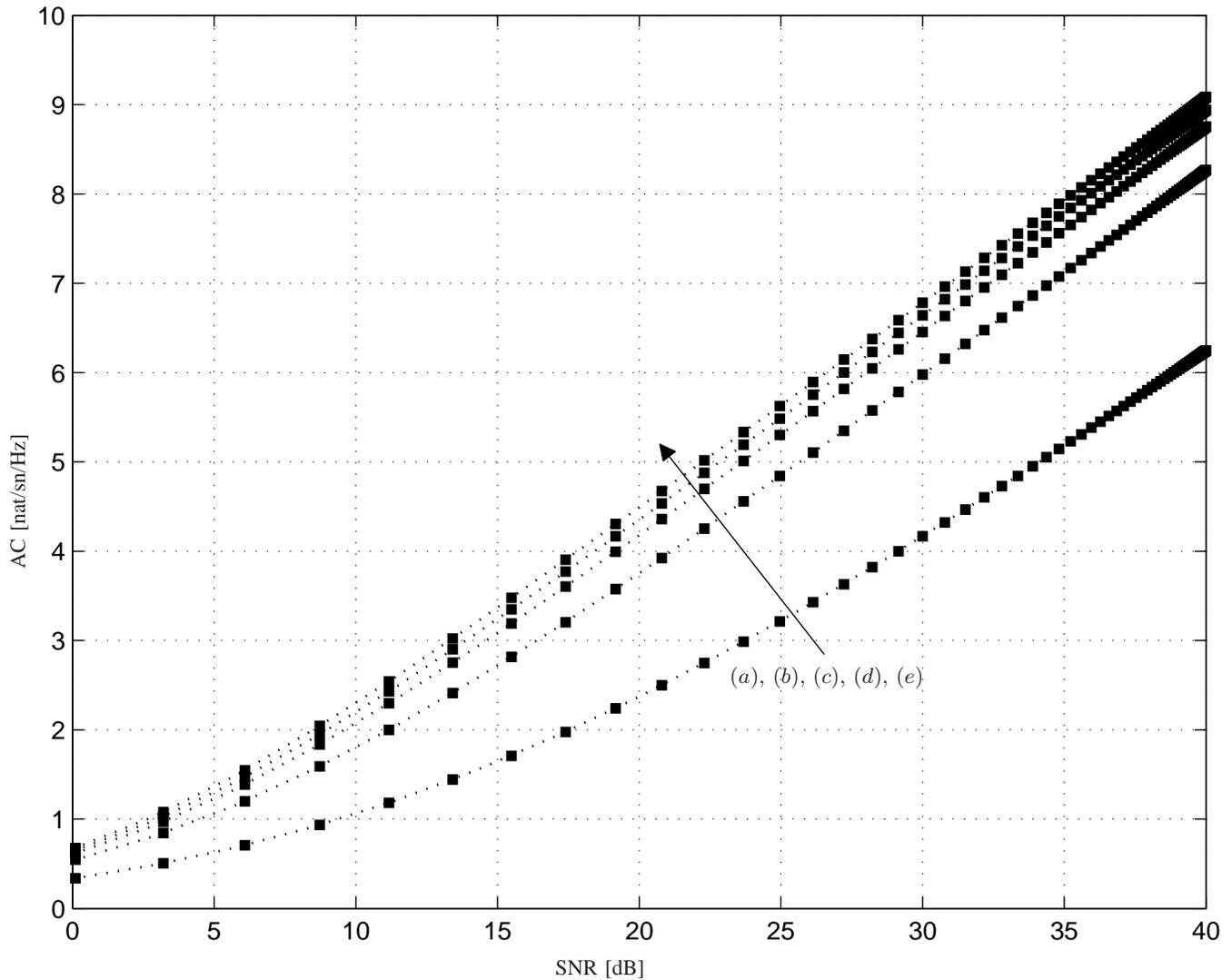

Fig. 3. AC of wireless communication system over generalized Gamma fading channel with the parameters: (a) $m_\ell = 2$, $\xi_\ell = 0.25$, (b) $m_\ell = 2$, $\xi_\ell = 0.5$, (c) $m_\ell = 2$, $\xi_\ell = 0.75$, (d) $m_\ell = 2$, $\xi_\ell = 1.0$ and (e) $m_\ell = 2$, $\xi_\ell = 1.5$. The number of samples is chosen as $10^6$ in the computer-based simulations.

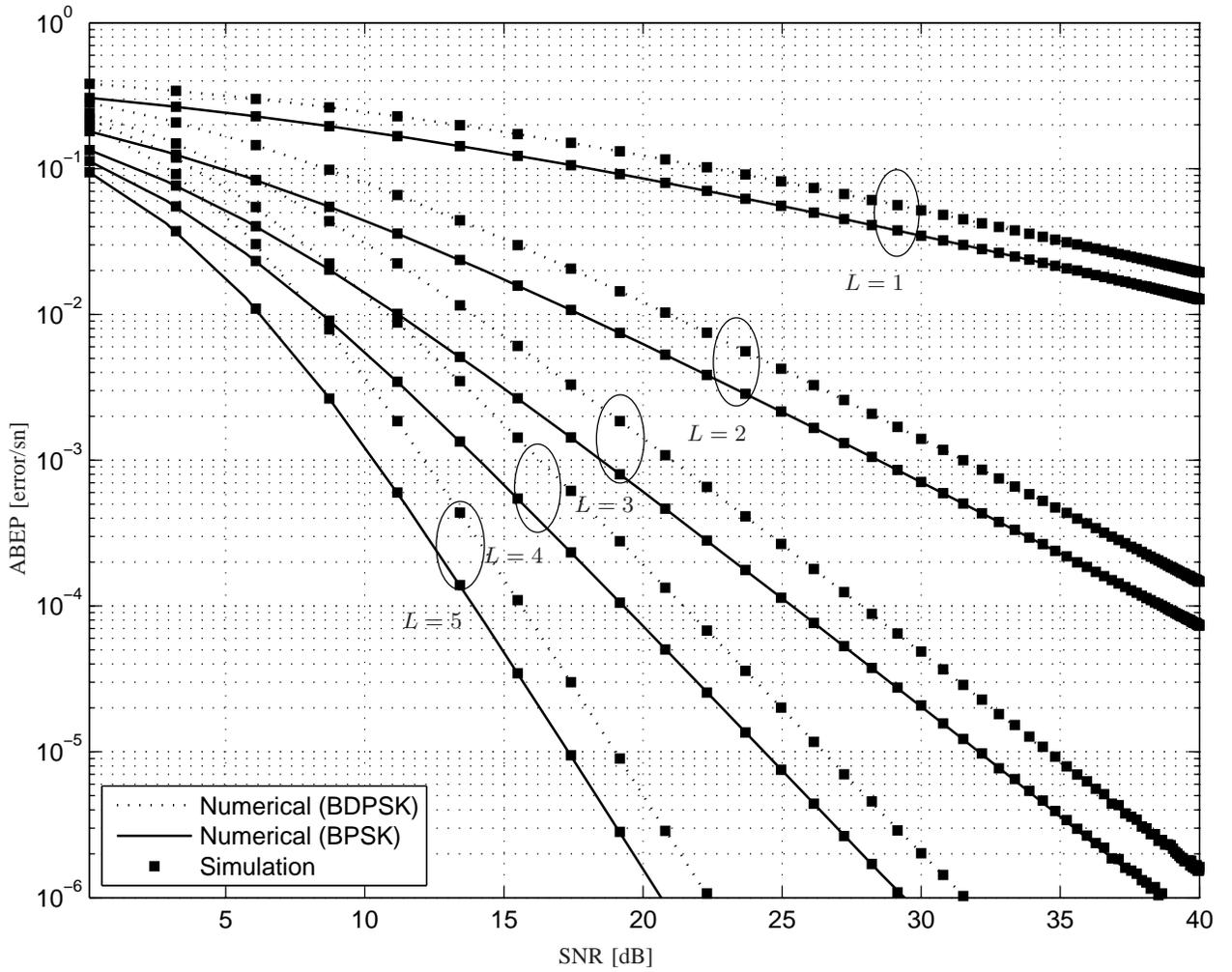

Fig. 4. ABEP performance comparison of BPSK and BDPSK binary modulations for $L$-branch MRC receiver over generalized Gamma fading channels with the parameters $m_\ell = 2$ and $\xi_\ell = 0.25$. The number of samples is chosen as $10^7$ in the computer-based simulations.

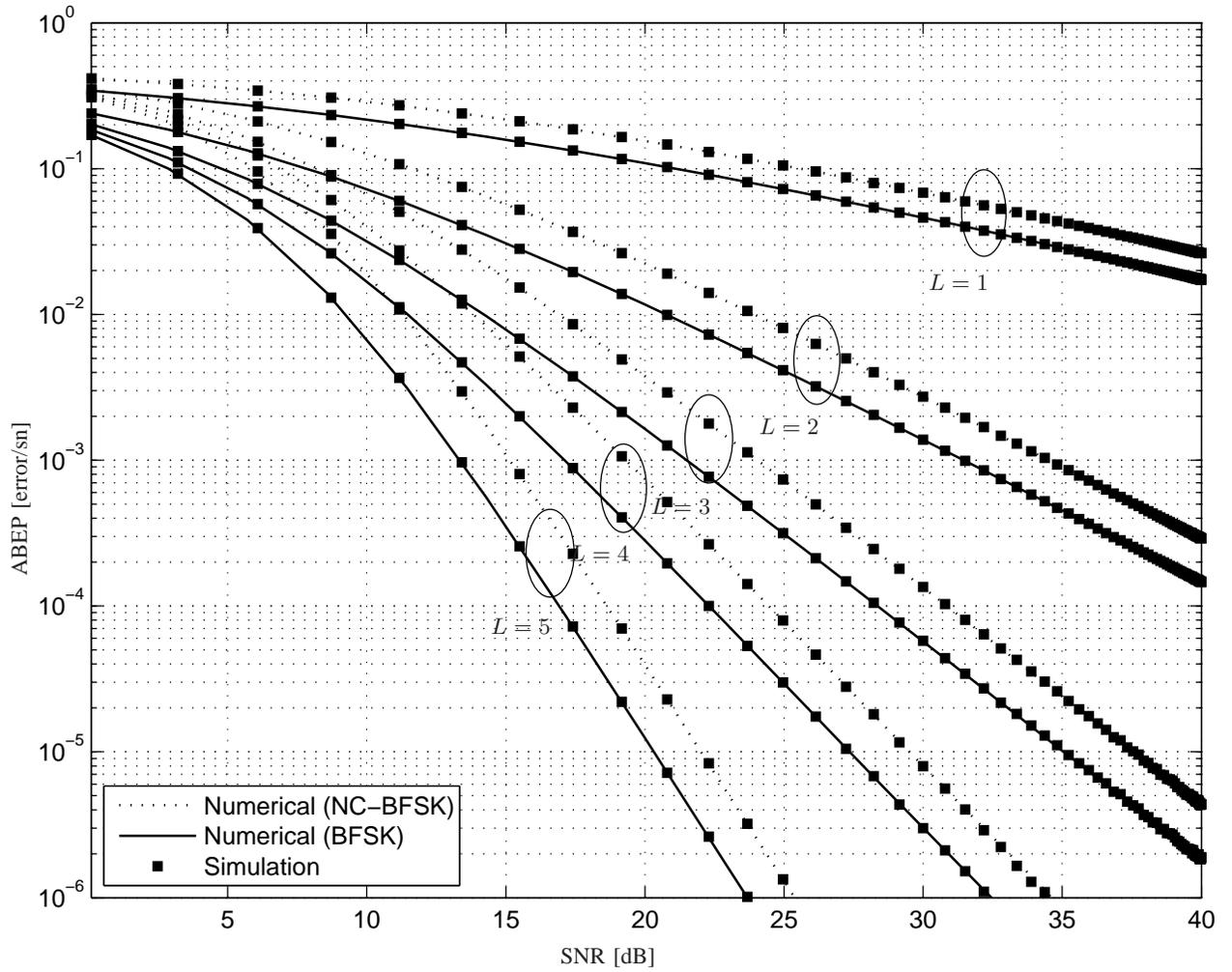

Fig. 5. ABEP performance comparison of BFSK and NC-FSK binary modulations for $L$-branch MRC receiver over generalized Gamma fading channels with the parameters $m_\ell = 2$ and $\xi_\ell = 0.25$. The number of samples is chosen as $10^7$ in the computer-based simulations.

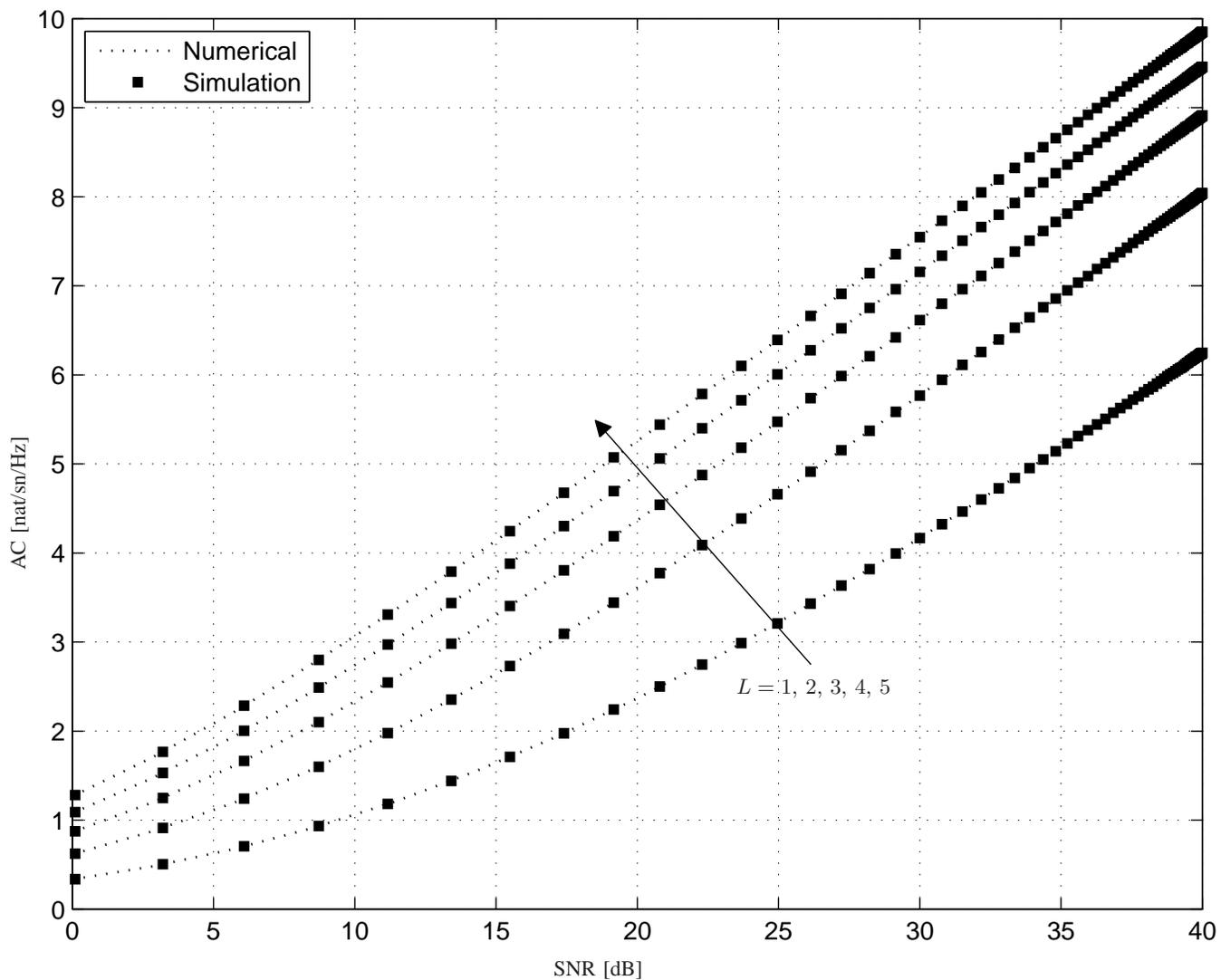

Fig. 6. AC of wireless communication systems with an $L$-branch MRC receiver over generalized Gamma fading channel with the parameters $m_\ell = 2$, $\xi_\ell = 0.25$. The number of samples is chosen as $10^6$ in the computer-based simulations.